\documentclass[
  journal=pasa,
  manuscript=article-type,
  year=2025,
  volume=37,
]{cup-journal}

\usepackage[nopatch]{microtype}
\usepackage{booktabs}
\usepackage{graphicx}	
\usepackage{amsmath}	
\usepackage{amssymb}
\usepackage{mathtools, nccmath}
\usepackage{subcaption}
\usepackage{comment}
\usepackage{hyperref}
\usepackage{cuted}

\graphicspath{{./}{figures/}}
\title{Polarisation Performance of Offset Phase Antennas: A Study for FARSIDE}

\author{Nivedita Mahesh}
\affiliation{Division of Physics, Mathematics, and Astronomy, California Institute of Technology, Pasadena, California 91125, USA}
\alsoaffiliation{School of Earth and Space Exploration, 
Arizona State University, Tempe, Arizona, USA}
\email[Nivedita Mahesh]{nmahesh@caltech.edu}

\author{Judd D. Bowman}
\affiliation{School of Earth and Space Exploration, 
Arizona State University, Tempe, Arizona, USA}
\author{Bharat K. Gehlot}
\affiliation{School of Earth and Space Exploration, 
Arizona State University, Tempe, Arizona, USA}
\author{Daniel C. Jacobs}
\affiliation{School of Earth and Space Exploration, 
Arizona State University, Tempe, Arizona, USA}


\begin{document}

\begin{abstract}
Several radio telescopes have been planned or proposed to be deployed on the Lunar farside in the coming years. These will observe the unexplored ultra-long wavelengths of the electromagnetic spectrum from the lunar farside's unique radio-quiet and ionosphere-free environment. One such lunar radio array is the NASA-funded concept - the Farside Array for Radio Science Investigations of the Dark Ages and Exoplanets (FARSIDE).  FARSIDE will operate over 100~kHz to 40~MHz with 128 spatially non-co-located orthogonal pairs of antenna nodes distributed over a $12\times12$~km area in a four-arm spiral configuration. Being on the lunar farside, this radio interferometer will be deployed by tele-operated rovers. The rover deployment mode could lead to a phase offset between each of the two orthogonally polarised antenna elements in the array, which are typically co-located. In this paper, we quantify the effects of such antenna phase offsets on the polarisation response and imaging performance of the lunar radio array. Modelling and analysing the FARSIDE dipole beams with and without offset, we find the latter leads to additional leakages into Stokes U and V corresponding to Muller matrix terms of M$_{2(0,1,2,3)}$ and M$_{3(0,1,2,3)}$. Using a custom simulation pipeline to incorporate all four Stokes beams of spatially co-located and non-co-located dipoles, we produce visibilities and simulated images for the GLEAM (GaLactic and Extragalactic All-sky MWA) sky model through the FARSIDE array. We find that for a pure Stokes I input sky, the output image maximum Stokes V/I flux ratio for the offset case has increased to $2.5\%$ versus $0.05\%$ for the co-located case. The additional Stokes V needs to be corrected since the detection of Electron Cyclotron Maser (ECM) emissions from exoplanets requires high-fidelity Stokes V measurements.
\end{abstract}

\section{Introduction} \label{sec:intro}
We are in a paradigm-shifting decade for radio astronomy with growing interest in expanding the observational window to frequencies below 30~MHz. This range of the low-frequency radio spectrum has information from two of the most significant areas of astrophysics: understanding our Universe's early history during the cosmic Dark Ages and characterizing exoplanets. However, to date, the sky below 30~MHz remains largely unexplored.

The Dark Ages is the period between the last scattering of the cosmic microwave background (CMB) photons and the appearance of the first luminous sources, spanning approximately a hundred million years ($1090 < z \lesssim 30$). Observing this period could potentially probe the primordial matter power spectrum over a vast three-dimensional volume. 
This would provide details on linear structure formation---how quantum fluctuations seeded the gravitational collapse and growth of structure in the Universe \citep{munoz2015}---and offer a test-bed for the standard cosmological model without the complication of highly non-linear baryonic effects. Observations from this epoch could access more Fourier modes in the density field than the CMB, leading to enhanced constraints on the masses of neutrinos and their hierarchy \citep{Mao2008} and on the imprints of primordial gravitational waves to reveal the complexity and energy scale of cosmic inflation \citep{2018ansari}.  Any departures from the well-constrained predictions of the standard physics would provide new insights into the formation of structure, potentially into the nature of dark matter \citep{Slatyer2013}, early dark energy \citep{Hill2018}, or exotic physics \citep{Clark2018}.  The lack of luminous astrophysical sources during the Dark Ages makes the 21-cm signal due to the spin-flip transition of neutral hydrogen---the most ubiquitous matter present---the only potential probe of this significant epoch in the Universe's history \citep{loeb2012}. The 21-cm signal from the Dark Ages is redshifted to wavelengths of 6-200~m today, observable in the 1-50~MHz radio frequency band.


In the same band, all magnetized planets in our solar system produce bright, highly circularly polarized, coherent emission. This emission originates predominantly from the polar regions of the magnetic field and is attributed to electron cyclotron maser instability \citep{zarka1998, ergun2000}. This radio emission is produced at the electron gyrofrequency, thus directly diagnosing the magnetic field strength in planetary magnetospheres. This is how the presence and strength of the Jovian magnetic field was first inferred \citep{Burke1955}. Planetary magnetospheres potentially play a role in the composition and retention of planetary atmospheres and may be a crucial ingredient for planetary habitability \citep{patsourakos2017}. This radio emission is, therefore, an important potential probe in understanding the role of magnetic fields in the habitability of exoplanets.

Detecting similar radio emissions from exoplanets around nearby stars will narrow the search for habitable planets beyond our Solar System.
 Exoplanet radio emissions are expected to have low flux densities, with the maximum occurring at low frequencies based on scaling models that use Jovian decametric emission. The average power from Jovian decametric emission is about $2.1\times10^{11}$~W with a spectrum that peaks at 22~MHz and has an upper cut-off frequency at about 40~MHz \citep{zarka2004}. Thus, to search for habitable planets beyond our Solar System, a radio array must be sensitive to frequencies below $\approx$ 40~MHz. We use the Jovian scaling to look for Jupiter-like exoplanets since observations of radio auroral emissions from other Solar System planets have shown that terrestrial ones have lower magnetic fields and lower flux densities of radio emissions. For example, the peak radio energy from Earth's Auroral Kilometric Radiation is two magnitudes lower than that of Jupiter's with a maximum of $10^{9}$~W observed at $\approx 119 - 500 kHz$. 
  The electron cyclotron maser (ECM) emission is produced primarily due to the interaction of solar winds with the planet's magnetosphere. The ECMs are known to be 100$\%$ circularly polarized, pulsed with a narrow duty cycle ($<10\%$), and with a typical rotation period of 2-3 hours \citep{hallinan2007,hallinan2008, berger2009}. In the event of such pulsed emission, 
 good sampling of the rotational and/or orbital period is required to ensure detection. In addition, the solar winds and coronal mass ejections needed to feed the auroral activity can be sporadic. This necessitates a low-frequency array continuously surveying multiple regions and multiple systems of the sky all at once, i.e., a pan-optic view of the sky.
Low-frequency radio astronomy can reliably be performed only from space. At frequencies below 40~MHz, there is heavy contamination on the Earth's surface by anthropogenic radio frequency interference (RFI), corruption by the ionosphere via its absorption and emission that scales with frequency as $\nu^{-2}$ and becomes dominant below $\approx$30~MHz \citep{vendantham2015, Rogers2015}, and refraction and scintillation by the ionosphere and solar wind plasma \citep{liu2011}. 

\subsection{Low-Frequency Facilities}

Only a few ground-based radio astronomy facilities have operated below 40~MHz. One of them is the Ukrainian T-shaped Radio Telescope, model-2 (UTR-2), which observes in the band $\approx$8-34~MHz \citep{Braude1978}. UTR-2 has produced several significant results in the field of low-frequency radio astronomy-related to antenna systems, equipment, and observational methodology and is the only ground-based instrument that has detected signals from Saturn lightning discharges. However, high-sensitivity pilot searches for radio transients from exoplanets and magnetars have been limited by ionospheric interference and RFI \citep{konovalenko2016}.  Decametric radio telescopes were also operated in the United States at Clark Lake \citep{Erickson1982} with a large effective collecting area and wide bandwidth (15 - 150~MHz). Primary beam and sidelobe confusion limited the effective sensitivity, resulting in the best frequency operation between 25-75~MHz. In France at Nançay, the Nançay Decameter Array (NDA) operated from 10-70~MHz \citep{Boischot1980}. The NDA, with its circular polarization-sensitive spiral antennas, has produced unique and continuous measurements that formed the basis of numerous solar and jovian studies. However, the NDA is not an interferometer but a single-phased array and cannot create images and only produces a single spectrum from the sky position observed \citep{lamy2018}. Some observations below 30~MHz were made from ground facilities at Tasmania (2-18 MHz) in the south and Canada in the north (10~MHz and 22~MHz) \citep{Reber1994,bridle1967, Cane1978, Roger1999}. Most of these efforts, at best, produced very low-resolution maps with limited observations in time and bandwidth due in large part to complications by ionospheric transmission.

Newer ground-based radio telescopes are designed to explore the sky at the frequencies of interest, including the LOw-Frequency ARray (LOFAR) in the Netherlands, the Owens Valley Radio Observatory-Long Wavelength Array  (OVRO-LWA)in California and the Array of Long Baseline Antennas for Taking Radio Observations from the Sub-Antartic (ALBATROS). LOFAR's low-band array is designed for frequencies between 10-90~MHz, but at frequencies below 30~MHz, the quality of scientific data is reduced due to variations in the total electron content of the ionosphere, which is extremely difficult to calibrate over long baselines (more than a few kilometers). This limits the operational bandwidth to 30-90~MHz for its key science projects \citep{gehlot2018}. Analysis of 31~hr of observations from  OVRO-LWA between 27-84~MHz for radio transients showed that many false detections below 40~MHz were primarily due to scintillation of sources caused by the ionosphere \citep{Anderson2019}. The initial pathfinders for ALBATROS at Marion Island in the Southern Indian Ocean and at the McGill Arctic Research Station on Axel Heiberg Island showed reasonable and repeatable sky fringes down to $\approx$ 10 MHz \citep{Chiang_2020}. ALBATROS relies on sites at polar or near-polar latitudes since such locations generally have lower ionospheric plasma frequency cutoffs than elsewhere on Earth \citep{Bilitza2018}. 

\subsection{Observations from Space}

Some space missions, such as WIND and Cassini, have carried low-frequency radio payloads. The data collected by these satellites showed that the Earth has strong natural radio emission at frequencies between $\approx$ 50 - 800~kHz (wavelengths between 6-0.3~km) called the Auroral Kilometric Radiation (AKR) \citep{zhao2019}.  Terrestrial transmitters observed from space are also strong, even with ionospheric attenuation reducing their propagation into space. Any space-based astronomical observatory in the inner Solar System, therefore, must find solutions to mitigate AKR and terrestrial RFI.

The lunar farside is one of the few truly pristine, radio-quiet platforms in the inner Solar System.  Since it is always facing away from Earth, the Moon itself acts as a shield to block terrestrial RFI and AKR.  In addition, excess system noise ($\leq$1~MHz) produced by electrons in the solar wind interacting with radio antennas is also reduced by the lunar wake cavity, especially at lunar night \citep{farrell1998}. Thus, the environment of the lunar farside is uniquely positioned for astronomical instruments to study magnetospheres of exoplanets and probe the very early Universe via low-frequency radio observations.

Recent and planned lunar missions are tapping into the advantages of the Moon's farside for low-frequency radio observations. China's Chang'e 4 successfully landed on the farside of the Moon on January 3, 2019 \citep{change}. This mission also deployed the Netherlands China Low-Frequency Explorer (NCLE) instrument on the orbiter, which is the only current radio instrument to study the unexplored regime of 80~kHz to 80~MHz from the lunar farside \citep{Vecchio2021}. It is the first receiver to enter the Lunar far side radio environment since NASA's Radio Astronomy Explorer-2 (RAE) in 1972; however, no results were reported. 
The Discovering the Sky at the Longest Wavelengths (DSL) mission concept \citep{chen2019} plans a constellation of 15 micro-satellites circling the Moon on nearly-identical orbits to form a linear array for interferometric observations below 30~MHz. Although the sensitivity of such an array is insufficient to detect the fluctuating 21cm signal from the Dark Ages or exoplanet auroral emission, it will make a practical first step by mapping Galactic foregrounds. These missions and experiments aim to demonstrate the value of space-based low-frequency radio astronomy for opening the parameter space of discovery and complementing ground-based efforts \citep{koopmans2021, burns2021nasa}.  They will build on the Sun Radio Interferometer Space Experiment (SunRISE) that is under development to study energetic particle acceleration at Coronal Mass Ejections (CMEs) by making the first spatially resolved observations of coherent Type II and III radio bursts from the Sun below 25~MHz. SunRISE will be a constellation of six spacecraft flying in a 10-km diameter formation in approximately geostationary orbits.  SunRISE will be the first imaging radio interferometer in space \citep{Kasper2018}.

A new era of low-frequency radio exploration of the Universe is planned from the Moon in conjunction with NASA's ongoing Artemis program\footnote{https://www.nasa.gov/specials/artemis/}. Specifically, NASA's Commercial Lunar Payload Services (CLPS) initiative allows for the rapid delivery of Lunar payloads for science experiments, technology testing, and exploration. CLPS missions have commenced since the beginning of 2024, with Astrobotic demonstrating a successful launch. Following that, Intuitive Machines carrying the first ever low-frequency instrument, Radio wave Observations at the Lunar Surface of the photoElectron Sheath (ROLSES) \citep{burns2021nasa}, landed on the moon. Unfortunately, the lander tipped over its side with the solar panels not facing the sun, and the mission had to end after 8 hours. But two of the four 2.5~m monopoles were deployed, and the group at CU Boulder is currently analysing eight hours of ROLSES data collected.  
The second planned NASA radio experiment with the CLPS program is the Lunar Surface Electromagnetics Experiment (LuSEE-Night) funded by the Department of Energy. Firefly Aerospace is scheduled to deliver it to the far side of the Moon mid-next year on its Ghost Lander. LuSEE Night, with its two 6~m long orthogonal dipoles and a 50~MHz Nyquist sampled base-band receiver system, is aimed at providing the first nighttime sky spectra in the band corresponding to the Universe's Dark Ages \citep{Bale_2023}. These CLPS radio science missions will prepare the way for the future low-frequency radio array on the surface of the Moon.


\subsection{FARSIDE}
\label{farside}
Ambitious efforts for large, science-capable instruments are underway, enabled by NASA's interest in returning humankind to the Moon. In November 2022, Artemis I, after its 1.4-million-mile mission beyond the Moon and back, successfully demonstrated the functioning and safety of the Orion Space Craft systems. The planned successive Artemis II mission, with its international and commercial partnerships, is now driving the development of Gateway, a crewed space station in a lunar halo orbit. The program is also advancing a variety of supporting technologies for lunar surface operations, including progressively more capable lunar landers, cold-tolerant electronics, and orbiting navigation and communication systems. Building on this anticipated infrastructure, the Farside Array for Radio Science Investigations of the Dark Ages and Exoplanets (FARSIDE) is a concept for a probe-class mission to place a low-frequency radio interferometer on the lunar farside surface. FARSIDE will take advantage of the Artemis and CLPS investments, which are expected to reach sufficient maturity by the mid-2020s to support a mission in the 2030s time frame. 

The notional architecture of FARSIDE consists of 128 dual polarization antennas spanning a $12\times12$~km area \citep{Burns2019a, McGarey2022}. Rovers will deploy and tether the array to a base station for central processing, power, and data transmission to the Lunar Gateway. The base station will house the X-engine and collect full cross-correlated visibilities every 60s to be transmitted to the Lunar Gateway every 24 hours. The visibilities will be calibrated and processed on Earth for science analysis. FARSIDE will provide the capability to image the entire visible sky each minute in 1400 channels spanning frequencies from 100~kHz to 40~MHz, extending down two orders of magnitude below bands accessible to ground-based radio astronomy.
 
FARSIDE will search for radio emissions from exoplanets and provide a testbed for demonstrating the technology needed to explore the Dark Ages through hydrogen cosmology.   To search for radio signatures of CMEs and exoplanetary radio emission, FARSIDE plans to observe 2000 stellar/planetary systems (approximately one system every 5~deg$^2$ on the sky) every 60 secs. This will enable near-continuous monitoring of the nearest stellar systems and achieve better sampling of the rotational phase of exoplanets that previous observations of hot Jupiters, typically a few hours in duration, have failed to achieve.  FARSIDE's very low-frequency range will make it considerably more sensitive to a range of planetary magnetic field strengths and stellar flares than Earth-based telescopes. 

To advance 21-cm Dark Ages observations, FARSIDE will employ two complementary approaches. First, the sky-averaged global signal or monopole will be observed by a spectrometer connected to individual dipole antennas. The observed brightness temperature is a gauge of the evolution of the neutral hydrogen density, along with the radio background and gas temperature. This mode will be supported by precision calibration via an orbiting beacon and sophisticated modelling of the instrument, lunar environment, and foregrounds. The second approach is interferometric measurements to constrain spatial fluctuations in the 21-cm Dark Ages signal.  In the long term, such observations can provide insights into linear structure formation, quantify departures from standard cosmological models \citep{Mao2008}, and shed light on the nature of dark matter \citep{slayter2016}. The planned array spacing and \textit{uv-}coverage will enable high-quality foreground imaging and pathfinder observations to constrain the 21-cm fluctuation power spectrum.  

\subsection{Offset Phase Antennas}
\label{offset}
Deploying FARSIDE on the lunar surface has added constraints compared to Earth-based radio arrays.  The reference plan for FARSIDE is to deploy the elements of the array, including antennas, receiver nodes, and cables for data relay and power supply with teleoperated rovers. Axel rovers \citep{nesnas2012} from the Jet Propulsion Laboratory will carry spools of tethers with dipole antenna elements embedded and unwind them on the lunar surface.  Thus, FARSIDE's array layout design is chosen to optimise the rover path and minimise the amount of material per rover. The two primary science cases described above warrant orthogonal dipoles for each antenna node to make dual-polarisation measurements. To accommodate the  FARSIDE design with many dipoles in sequence along a tether, the planned array layout will use two sequential dipoles to create each dual-polarisation antenna node.  The rover unwinding the tether will make a right-angle turn between the two dipoles so that the dipoles have orthogonal polarisation.  This deployment causes the phase centres of the two dipoles to be spatially offset, in contrast to most existing radio telescopes, and gives the deployed tether a stair step appearance (refer to the inset of Figure \ref{fig:array-layout-artist}).

Here, we study the imaging performance that results from these polarisation phase offsets in FARSIDE antenna nodes. We look at the direction-dependent effects of the primary beam and antenna phase centre offsets on the output polarisation images. The analysis presented in this work is generalised and can be extended to any antenna beam and array layout. Direction-dependent primary beams cause intermixing of polarisation components in any array with cross-polarised feeds.  Phase offset between antennas is seen in arrays like the 21CMA (also called PaST) and on the focal plane array of ASKAP and APERTIF, making the mathematical framework used in this analysis applicable to those arrays.

In section \ref{sec:array_design}, we describe the planned array layout for FARSIDE. We look at the effects of antenna spatial offsets using simple visualisations in section \ref{sec:offset_effect}. An overview of the mathematical formalism used for the polarisation analysis of the antenna offsets is covered in section \ref{sec:formulism}. In section \ref{sec:stokes_leakage}, we simulate as a function of frequency and quantify the direction-dependent polarisation leakages due to the dipole beam and offsets. We extend the analysis and quantify the polarisation leakages on simulated sky observations in section \ref{sec:effects_onsky}. In section \ref{sec:correction} we suggest a correction to improve the imaging performance of the array. Finally, we conclude with a few comments on future work.

\section{FARSIDE Design and Nominal Array Layout}
\label{sec:array_design}
\begin{figure}
\centering
\includegraphics[width=0.9\textwidth]{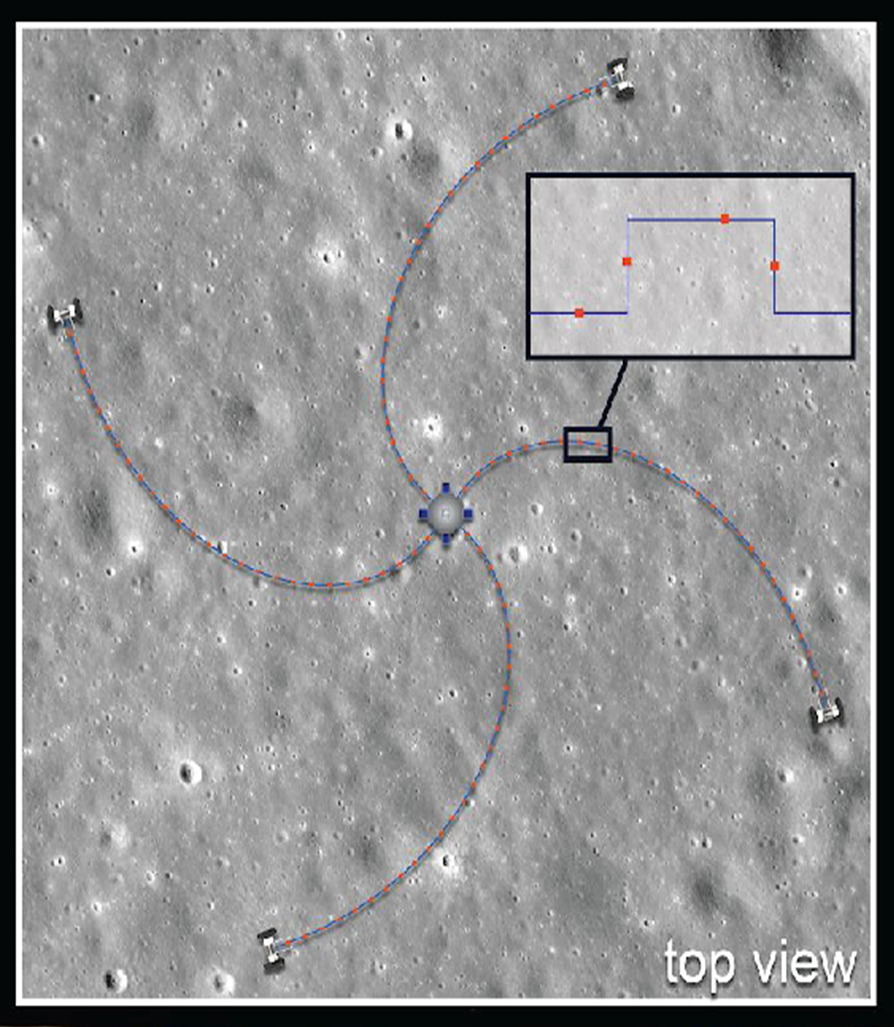}
\caption[FARSIDE Array Layout]{\label{fig:array-layout-artist} An artist's rendering of the four arm spiral configuration of the FARSIDE array on the lunar surface. At the centre of the array is the base station with the communication antenna, fuel tank, central processing unit with correlators and the main power supply. Each of the four spiral arms, will have 32 antenna nodes consisting of two dipoles and a receiver. Also shown are the four two-wheeled rovers that will deploy the tethers containing the antenna nodes. The inset image shows the path taken by the rover to lay out the embedded dipole antennas with the 90-degree bend at each antenna node. The phase centres of the dipoles are indicated by the red dots.} 
\end{figure}

\begin{figure}
\centering
\includegraphics[width=0.9\textwidth]{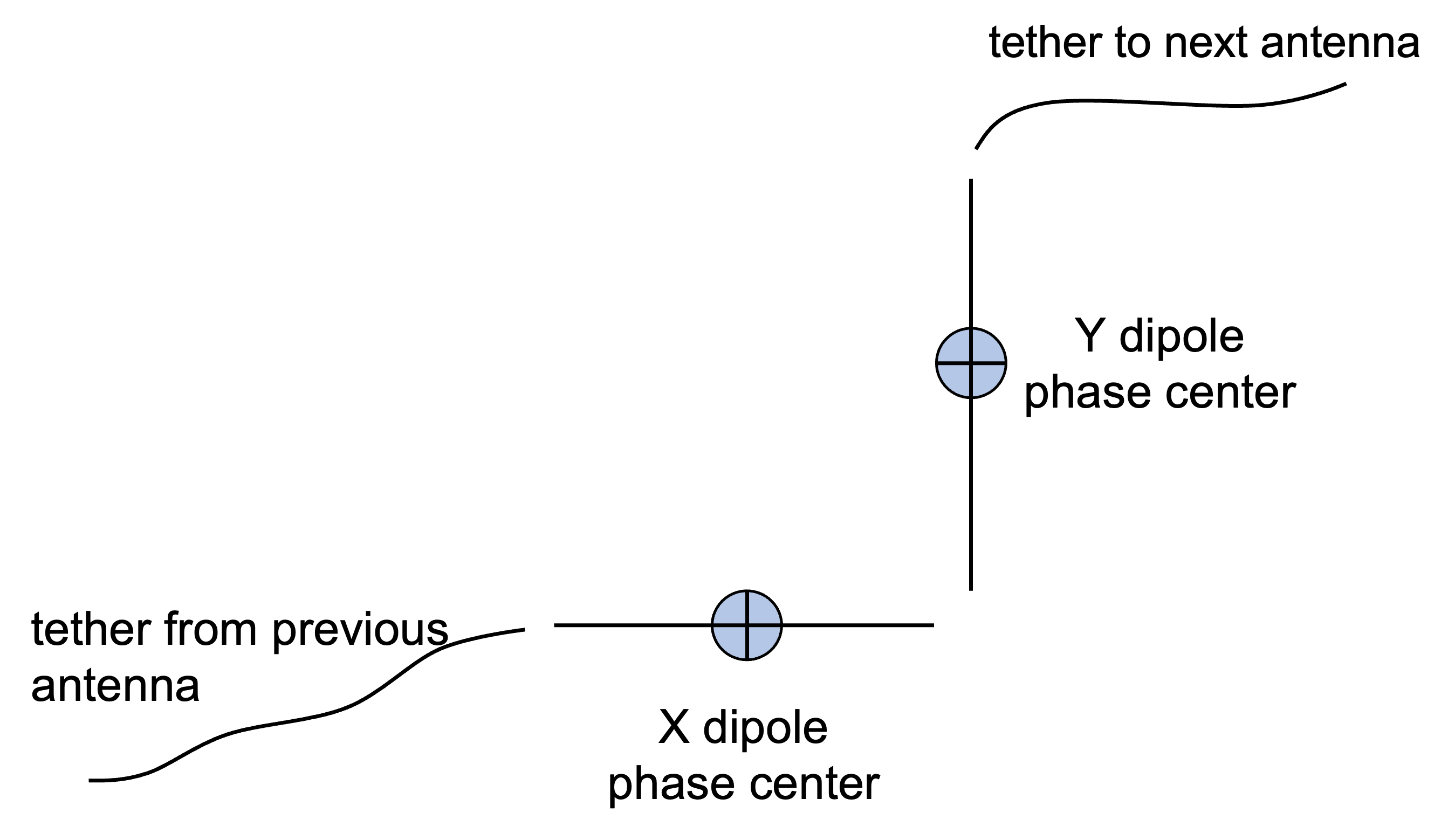}
\caption[FARSIDE Deployment Strategy]{\label{fig:antenna-node-deploy} A sketch detailing the deployment configuration for a single antenna node of FARSIDE. The tether from the previous antenna node leads up to one of the dipoles (X-dipole). Then, the rover turns 90${^\circ}$ and lays out the orthogonal dipole (Y-dipole), carrying the tether over to the location of the next node.  } 
\end{figure}

  As discussed in \autoref{farside}, the currently planned array layout of the FARSIDE instrument is a four-arm spiral with 32 pairs of spatially non co-located dipoles on each arm that are tethered to the central base station on the lander (as shown in \autoref{fig:array-layout-artist}). This array layout was arrived after analysing a few different configurations, including a four-petal, tighter spiral and fewer arm spirals \citep{Burns2019a}. This final array layout could be studied for further optimisation, including slight perturbations, but that would have to be balanced against the spool length and surface variation constraints. The central base station provides communication, data relay, and power during deployment as well as for operations via the tethers. The complete array and receiver nodes will be distributed over a $12\times12$~km area by four teleoperated, solar-powered, two-wheeled rovers that will deploy the antenna nodes over a single lunar day (14 Earth days) \citep{burns2021lunar}. To enable the desired science cases, the antenna and the array are designed to operate from 100~kHz to 40~MHz. The chosen length of the dipole is 100~m for the entire bandwidth of operation. These dipoles are too long to be deployed using the Spiral Tube and Actuator for Controlled Extension/Retraction (STACER), which have been used by numerous satellites for monopole and dipole antennas and in science investigations to study the solar wind. Based on the deployment strategy discussed in \autoref{offset}, the planned design is to embed the dipoles sequentially in the tethers that connect the individual nodes \citep{McGarey2022}.  In this scenario, the rover will deploy the tether with 90$^\circ$ bends along its path between each antenna location to ensure that the two embedded dipoles are aligned to orthogonal polarizations. This will lead to a spatial offset between the phase centres of the dipoles and the planned offset of 50~m along the X and Y directions (\autoref{fig:antenna-node-deploy}). This design provides for dual-polarisation measurements and full Stokes measurement of the electric field. 
\section{Spatial Offset in Interferometers}
\label{sec:offset_effect}
\begin{figure}
\centering
\includegraphics[width=0.85\textwidth]{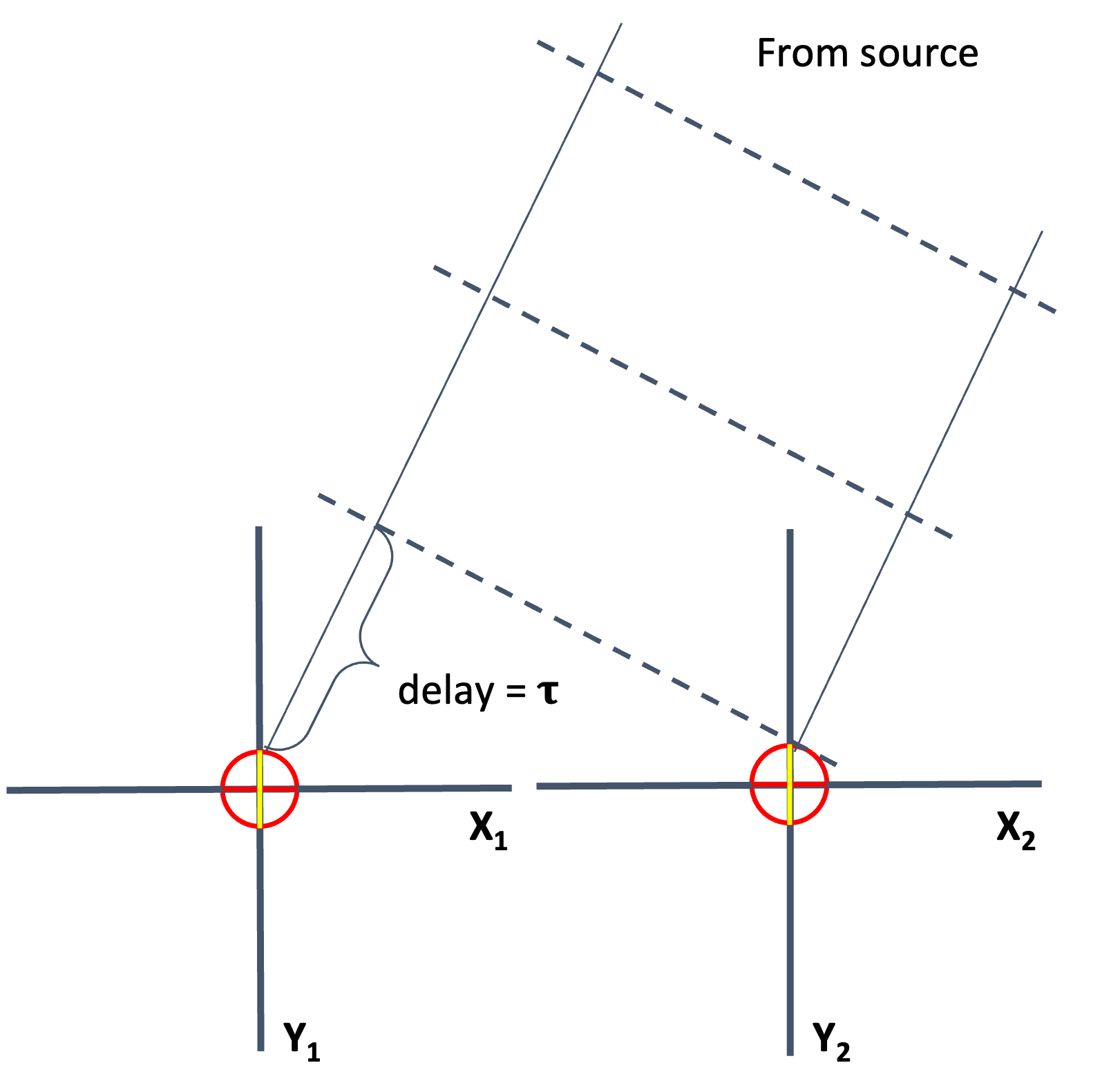}
\includegraphics[width=0.85\textwidth]{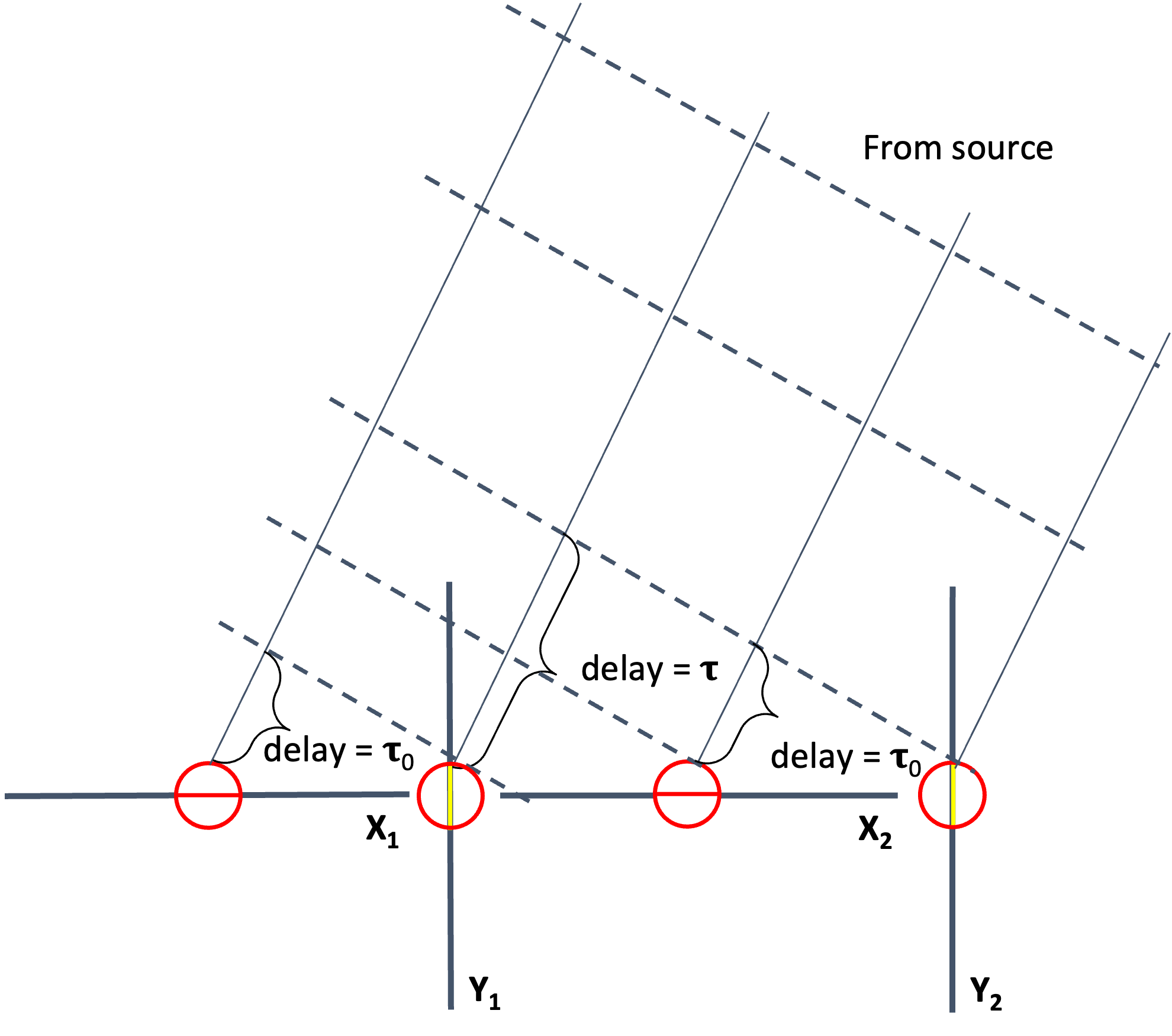}
\caption[Two-element Interferometer Example]{\label{fig:schematic_2element} A schematic highlighting the difference between the spatially co-located and non co-located dipoles in a 2 element interferometer. The offset between the phase centres results in an additional delay($\tau_o$) between the X and Y combinations of each antenna pair. Additional corrections are needed when cross-correlating data from different antennas.} 
\end{figure}

The offset between the phases of orthogonal dipoles of the same antenna leads to polarisation inter-mixing or leakages. In this work, we analyse the effects in detail and quantify them as impacts on the beam response and simulated un-deconvolved images. But in \autoref{sec:correction}, we also show that the phase offset effects can be compensated for with a simple correction.
We offer three ways of visualising and modelling the offset polarisation effects. First, in this section, we briefly describe a single baseline. Then, in  \autoref{sec:psf}, we see how the offsets can be accounted for as differences in the UV distributions of the X and Y polarisations.  In \autoref{sec:formulism} we will treat the offsets as internal properties of the antennas. 
We can gain intuition about the effects of offset orthogonal dipoles in the same antenna pair using ray tracing with a simple two element interferometer. In the top panel of \autoref{fig:schematic_2element}, the X and Y dipoles of each antenna node are spatially co-located. In this case, a single geometric delay correction is applied to both orthogonal dipoles to coherently observe towards any off-axis source/direction. Most current and next-generation radio interferometers such as HERA, LOFAR, MWA, LWA, GMRT, VLA  and SKA have spatially co-located dipoles. 

In the bottom panel of \autoref{fig:schematic_2element}, we introduce a spatial offset between the two orthogonal dipoles of each antenna. This is a simplified schematic showing a delay in a single dimension.    The spatial offset results in an additional delay of $\tau_o$ between the X and Y dipole of each antenna. Any baseline aligned along the offset direction will acquire this delay in addition to the usual geometric terms ($\tau$), for example Y2 to X1 will be $\tau + \tau_o$. 

\begin{figure}
\centering
\includegraphics[width=0.9\textwidth]{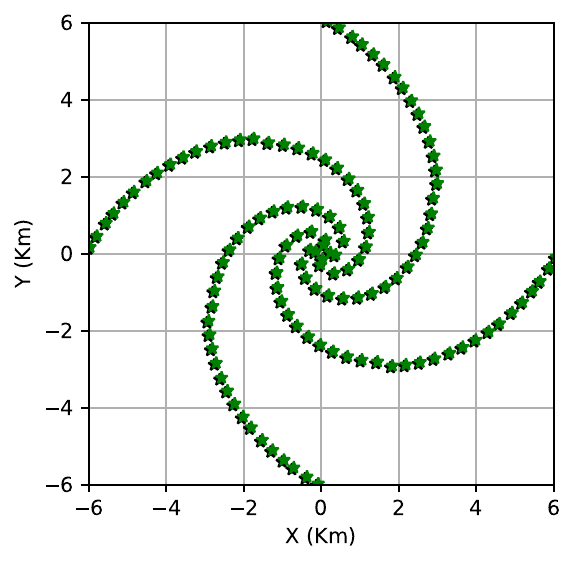}
\includegraphics[width=0.9\textwidth]{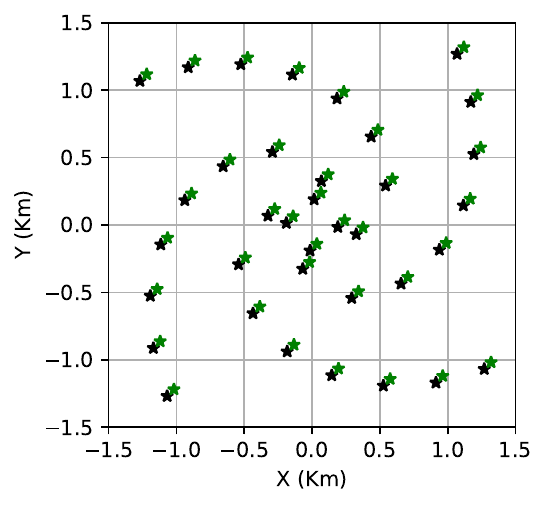}
\caption[FARSIDE Array Simulation]{\label{fig:spiral-array} Simulation of the dipole phase centres of the FARSIDE spiral arm array layout. Each arm has 32 pairs of dual-polarised dipoles. The green phase centres are offset from the black by 50~m in the X and Y directions. The top panel shows the top view of the complete layout spanning over 12~km in the X and Y extents. The bottom panel shows the inner $3\times3$~km of the layout and a closer look at the offsets between the X- and Y- dipoles in each antenna node.} 
\end{figure}

\subsection{Quantifying Geometric Delay for FARSIDE}
\label{sec:psf}
 To see if the offsets of the two polarisations will have an effect on the \textit{uv}-coverage and resulting Point Spread Functions (PSFs), we simulate visibilities using FARSIDE antenna positions (phase centres) of the spiral configuration, shown in \autoref{fig:spiral-array}. The top panel of the Figure shows the complete extent of the array spanning from -6~to 6~km in the X and Y spatial directions. The two colours, black and green, indicate the two polarisations of the array. The bottom plot zooms into the array's centre to highlight the offsets in the antenna positions. Following the notional design of FARSIDE discussed in \autoref{sec:array_design}, we take the offset between the orthogonal polarisations to be 50~m in both the X and Y directions. This is the minimum offset required to fit two dipoles, each with a half-arm length of 50~m, sequentially. For all the analyses in this paper, we assume the ideal offsets between the X and Y polarisation to be $\Delta x=50$~m and $\Delta y=50$~m.

\begin{figure*}
\begin{subfigure}{.48\textwidth}
\includegraphics[width=\textwidth]{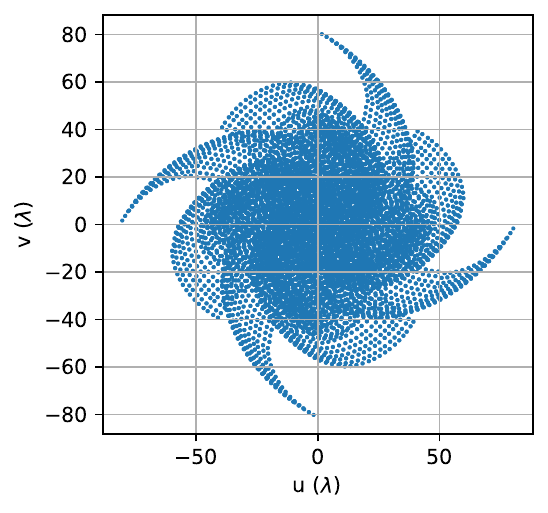}
\vspace*{-5ex}
\caption{\label{uv-xx}\textit{uv}-coverage for XX and YY baseline pairs.}
\end{subfigure}
\begin{subfigure}{.48\textwidth}
\includegraphics[width=\textwidth]{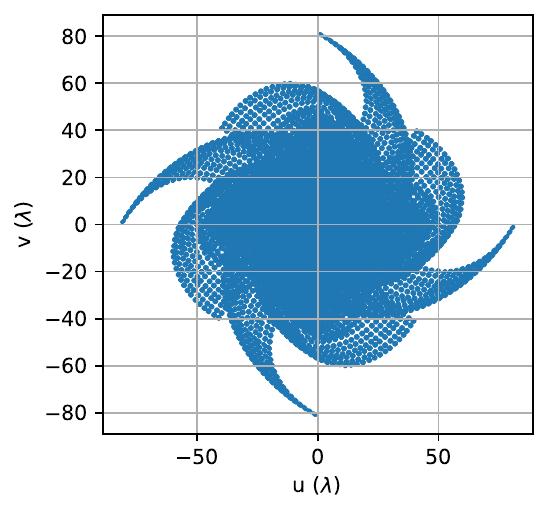}
\vspace*{-5ex}
\caption{\label{uv-xy}\textit{uv}-coverage for XY baseline pairs.}
\end{subfigure}
\newline
\begin{subfigure}{.48\textwidth}
\includegraphics[width=\textwidth]{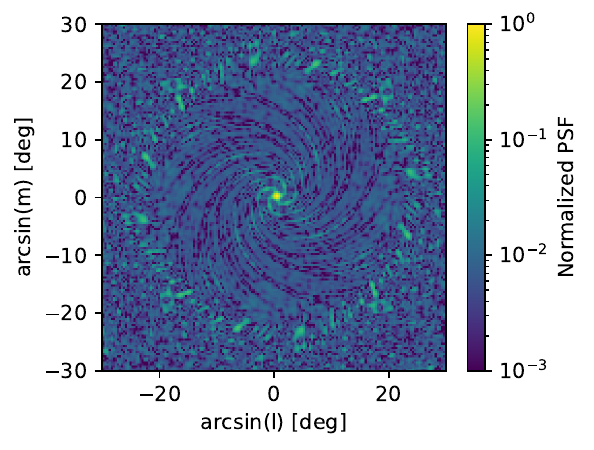}
\vspace*{-5ex}
\caption{\label{psf-xx}PSF of the XX/YY baseline pairs}
\end{subfigure}
\begin{subfigure}{.48\textwidth}
\includegraphics[width=\textwidth]{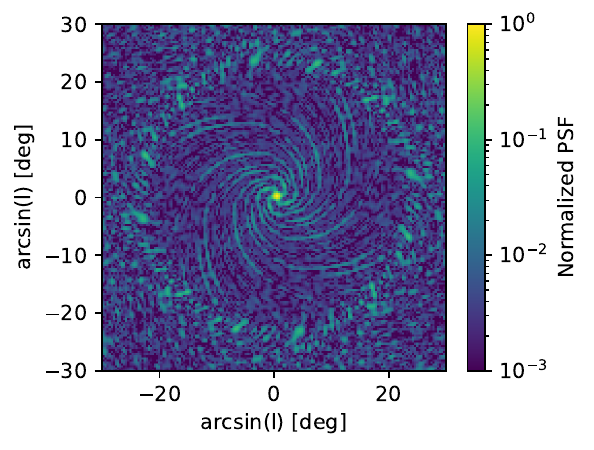}
\vspace*{-5ex}
\caption{\label{psf-xy} PSF of the XY baseline pairs}
\end{subfigure}
\newline
\begin{subfigure}{\textwidth}
\centering
\includegraphics[width=0.458\textwidth]{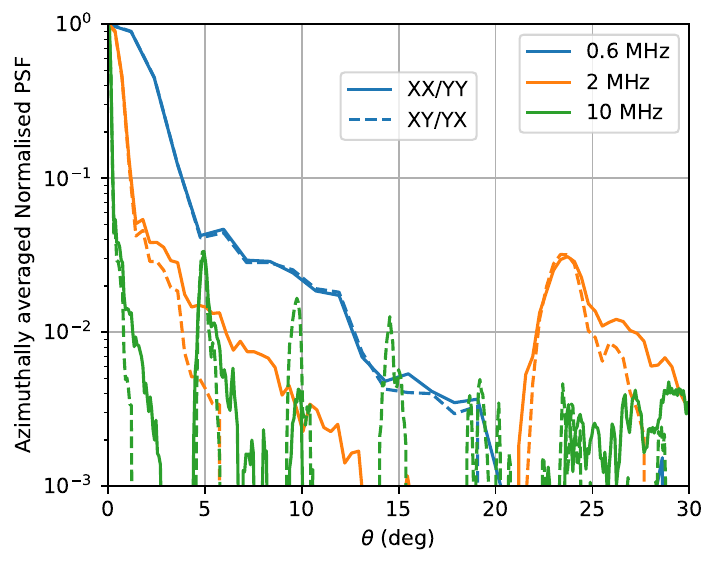}
\vspace*{-3ex}
\caption{\label{1dpsf}Azimuthally averaged PSF} 
\end{subfigure}
\caption[\footnotesize \textit{uv-}coverage and PSF of FARSIDE]{[a,b] Snapshot \textit{uv-}coverage at 2~MHz of the four arm spiral array layout for zenith pointing. [a] \textit{uv-}coverage for the XX and YY baselines and [b] shows \textit{uv-}sampling for the XY baselines of the antenna pairs. [c,d] Normalised 2D Point Spread Functions (PSF) of the FARSIDE spiral arm layout with and without offset. [e] Azimuthally-averaged PSF versus elevation angle for the XX/YY and XY sets of baselines of the FARSIDE spiral arm layout plotted for three characteristic frequencies within the operating bandwidth.}  
\end{figure*}
 The offset causes the sets of XY and YX baselines to fill different regions in visibility space compared to the corresponding XX and YY polarisations. \autoref{uv-xx} and \autoref{uv-xy} show the simulated \textit{uv-}coverages for XX/YY and XY/YX sets of baselines, respectively, at 2~MHz.  Given the maximum baseline of the array is 12~km (\autoref{fig:spiral-array}), the maximum \textit{uv-}bin is $\pm$80$\lambda$ (\autoref{uv-xx} and \autoref{uv-xy}). The perturbed array (with offset) will fill in the \textit{uv-}space better than the XX/YY case. This in turn results in a better PSF for the XY baselines as shown at the bottom of \autoref{1dpsf}, where the sidelobes are lower with deeper nulls for the XY case baseline at zenith angles greater than the resolution of the array at each frequency.

 Although offset polarisation effectively doubles the UV coverage density and improves cross-polarised PSF considerably, this model does not take into account the primary beam, which will affect the accuracy of polarisation measurements.

\section{Polarisation Leakage Due to the Beam}
\label{sec:formulism}

The beam of the dipole antenna has a pronounced gain pattern across the sky. When the two polarisation feeds have unequal gains, even an unpolarized sky appears polarised. In the case of offset phase centres, this polarised sky will also appear to have additional phase offsets. Through simulation and analysis, we will show that these phase offsets will vary with frequency and antenna offset errors. 


 For the case of an ideal antenna placement with no errors in deployment, we can quantify the effect of constant offset on the polarisation leakage of FARSIDE by cross-multiplying the beams of individual dual-polarisation antenna nodes. To estimate the total polarisation leakage caused by the FARSIDE array, we will propagate the obtained single-node polarisation beams to the interferometer pipeline by convolving the \textit{uv-}coverage (\autoref{sec:effects_onsky}).
 
\subsection{Review of the RIME formulation}

Consider a simple interferometer with just two antennas represented by $p$ and $q$, and each of these antennas has two orthogonal feeds ($x$ and $y$) sensitive to the two polarisations of the incoming wave. 
The two orthogonal feeds of each antenna produce voltages proportional to the sum of the electric fields ($e_{x}$
and $e_{y}$) for each polarisation at the location of the antenna. 
These voltages are products of the antenna properties and sky electric fields. Expressing the antenna properties in a Jones matrix ($J_p$)
for each antenna $p$, 
we can write the voltages ($\overline{v}_p$) at a particular antenna as follows:

\begin{equation}
\begin{split}
       \overline{v}_p & =  \begin{bmatrix}
        v_{px} \\
        v_{py}
    \end{bmatrix} = J_p \begin{bmatrix}
        e_{x} \\
        e_{y}
        \end{bmatrix} .
\end{split}
\end{equation} 
 The beam of the dipole antenna has a pronounced gain pattern across the sky. Even an unpolarised sky appears polarised when the two polarisation feeds have unequal gains. In the case of offset phase centres, this polarised sky will also have additional phase offsets. Through simulation and analysis, we will show that these phase offsets will vary with frequency and antenna offset errors.

The voltages from two antennas can be cross-correlated to produce a visibility matrix that depends on the sky coherency matrix \citep{born_wolf_bhatia_clemmow_gabor_stokes_taylor_wayman_wilcock_1999}. For polarisation studies, as in our case, it is more convenient to express this in terms of the visibility vector and, equivalently, the coherency vector \citep{hamaker1996}.  This is obtained by taking the outer product or the Kronecker product of the two input voltage vectors:
\begin{equation}
\begin{split}
        V_{pq} & = <\overline{v}_p \otimes \overline{v}_q^* > \\
        & =\langle J_p \begin{bmatrix}
             e_{x} \\
        e_{y}
        \end{bmatrix} \otimes  \left(J_q\begin{bmatrix}
             e_{x} \\
             e_{y} 
        \end{bmatrix}\right)^* \rangle , \\
\label{kroncker}
\end{split} 
\end{equation}
where $^*$ denotes element-by-element complex conjugate of the matrix and $<>$ indicates we calculate the average over time. Applying the property of the Kronecker product that holds true for any four matrices, $(A\otimes B) (C \otimes D) = AC \otimes BD$, allowing us to write:
\begin{equation*}
    \begin{split}
        V_{pq} = \langle\begin{bmatrix}
        v_{px}v_{qx}^* \\
        v_{px}v_{qy}^* \\
        v_{py}v_{qx}^*\\
        v_{py}v_{qy}^*
        \end{bmatrix} \rangle = (J_p\otimes J_q^*) \langle\begin{bmatrix}
             e_{x}e_{x}^* \\
        e_{x}e_{y}^* \\
        e_{y}e_{x}^*\\
        e_{y}e_{y}^*
        \end{bmatrix} \rangle
    \end{split}  .
\end{equation*}
For a single unresolved source, if we take into account the phase delay between the antennas $p$ and $q$, we can represent the above equation in terms of the \textit{source} coherency vector $\langle[ e_{x}e_{x}^* ~~e_{x}e_{y}^* ~~e_{y}e_{x}^* ~~ e_{y}e_{y}^* ]^T\rangle = \mathcal{E}_{sky}$ since the source is spatially incoherent. For this, we need the baseline vector between the two antennas in Cartesian coordinates represented by \textit{u,v,w} and as a function of wavelength; 
\begin{equation*}
    \begin{split}
        \langle\begin{bmatrix}
        V_{pq}
        \end{bmatrix} \rangle = (J_p\otimes J_q^*) \langle\begin{bmatrix}
        e_{x}e_{x}^* \\
        e_{x}e_{y}^* \\
        e_{y}e_{x}^*\\
        e_{y}e_{y}^*
        \end{bmatrix} \rangle \exp{(-2\pi i ( u l + v m + w n))},
    \end{split}
\end{equation*}
where \textit{l, m, n} are direction cosines that represent the coordinates of the source in the sky. To observe an extended region of the sky instead of a single source, we have to integrate the above equation over all the direction cosines of the sky (van Cittert Zernike theorem, \citet{thompson2001}):

\begin{equation}
    V_{pq} = \int \int (J_p\otimes J_q^*) \mathcal{E}_{\text{sky}}~ \exp{(-2\pi i ( u l + v m + w n))} \frac{dl dm}{n} .
\end{equation}
When imaging a small region ($\theta < 30^\circ$) of the sky or making the "flat-sky" approximation where \textit{w} = 0 or where $l^2 + m^2 << 1$ and the \textit{n} direction cosine ($n=\sqrt{1-l^2 - m^2}$) evaluates to $\approx 1$, the above equation simplifies to:
\begin{equation}
    V_{pq} = \int \int (J_p\otimes J_q^*) \mathcal{E}_{\text{sky}}~ \exp{(-2\pi i (u l + v m))} dl dm ~.
    \label{visilibity}
\end{equation}

We note that unless $J_p$ is both diagonal and has, at any given point on the sphere, equal diagonal elements, there will be mixing or “leaking” of different Stokes parameters together into each element of the visibility vector in a direction-dependent way \citep{geil2011, smirnov2011, nunhokee2017,asad2016}. We can expand $J_i$ into a product of matrices, each representing different antenna properties.  In our analysis, we look at the beam and offset properties and analyse if and how these cause intermixing of the various polarisation components. 

To conserve space, we will represent $\exp{(-2\pi i ( u l + v m))}$ as $K_{pq}$.
\subsection{Stokes Polarimeter and Muller Matrix}

In all our calculations till now, we have represented the sky coherency vector in the Cartesian frame. The Stokes frame will give us the details of the source's polarisation leaking into the polarisation components of the system. So, we will apply a coordinate transform to the above equation in the Cartesian system and transfer it to the Stokes system. We use the unitary transform matrix (S) on the linear operator J as: $S^{-1} (J_p\otimes J_q^*) S$ \citep{hamaker1996}. The transformation matrix (S) used here (Equation \ref{eqn:s}) multiplied by a scaling factor of $\frac{1}{\sqrt{2}}$ is unitary such that $(\frac{1}{\sqrt{2}}S)^{-1} = \frac{1}{\sqrt{2}}S^\dagger$ ( $^\dagger$ represents the complex conjugate transpose). The term $S^{-1}(J_p \otimes J_q^*)S$ is the defined as the Muller matrix ($M_{pq}$). It helps us quantify how the sky Stokes components are received by the XX, XY, YX, and YY components of the voltage vectors. And $\mathcal{E}^S_\text{sky} = S^{-1}\mathcal{E}_\text{sky}$ is the coherency vector of the sky in terms of the Stokes coordinate frame; $\mathcal{E}_{\text{sky}}^S = (I~Q~U~V)^T$. We apply the coordinate transformation to \autoref{visilibity} using these definitions:
\begin{equation}
    S = \begin{bmatrix}
        1 & 1 & 0 & 0\\
        0 &  0 & 1 & i \\
        0 & 0 & 1 & -i \\
        1 & -1 & 0 & 0\\
    \end{bmatrix} ,
    \label{eqn:s}
\end{equation}
\begin{equation}
\begin{split}
     S^{-1} V_{pq} = \int \int S^{-1}(J_p\otimes J_q^*)  S ~(S^{-1}\mathcal{E}_{\text{sky}}) K_{pq} dl dm~, \\
            \begin{bmatrix}
        1 & 0 & 0 & 1\\
        1 &  0 & 0 & -1 \\
        0 & 1 & 1 & 0 \\
        0 & -i & i & 0\\
    \end{bmatrix} \begin{bmatrix}
             v_{px}v_{qx}^* \\
        v_{px}v_{qy}^* \\
        v_{py}v_{qx}^*\\
        v_{py}v_{qy}^*
        \end{bmatrix} = \int \int S^{-1}(J_p\otimes J_q^*)  S ~\mathcal{E}^s_{\text{sky}} K_{pq} dl dm~, \\
        \end{split}
\end{equation}
\begin{equation}
    \begin{split}
    \begin{bmatrix}
          v_{px}v_{qx}^* + v_{py}v_{qy}^* \\
             v_{px}v_{qx}^* - v_{py}v_{qy}^*\\
             v_{px}v_{qy}^* + v_{py}v_{qx}^* \\
        -iv_{px}v_{qy}^* +i v_{py}v_{qx}^*       \end{bmatrix} =   \begin{bmatrix}
          \mathcal{V}_I \\
          \mathcal{V}_Q \\
          \mathcal{V}_U \\
          \mathcal{V}_V 
      \end{bmatrix} &= \int \int S^{-1}(J_p\otimes J_q^*)  S ~\mathcal{E}^s_{\text{sky}} K_{pq} dl dm \\
      & =\int \int M_{pq} ~\mathcal{E}^s_{\text{sky}} K_{pq} dl dm ~.
      \label{eqn:vis-final}
     \end{split}
\end{equation}

For example, to understand how all the Stokes components of the sky enter the measured or pseudo-Stokes ($\mathcal{V}$), we look at just the first component of the $\mathcal{V}$ vector, which is  

\begin{equation*}
\begin{split}
       \mathcal{V}_I = \int (M00_{pq}I +  M01_{pq}Q + & M02_{pq}U + M03_{pq}V) K_{pq} dl dm ~.
\end{split}
\end{equation*}

\subsection{Methods of Quantifying the 
Polarisation Performance of an Array}
\label{muller_steps}
We estimate and quantify the effects of the beam and dipole phase offsets by calculating the Muller matrices defined above. This calculation is carried out using Muller matrices for the representation of the Stokes leakages. The Muller matrices are calculated in the following manner:

First, we define the Jones matrix (\textit{J}) for the effects of the array to be analysed. For the co-located case, we only use the direction-dependent beam of antenna node \textit{p}. The electric beam patterns to determine $J = J_{beam}$ are obtained from the FEKO electromagnetic simulations. The simulation coordinates are represented by $\theta$, which corresponds to the elevation angle with zero at zenith, and $\phi$, which corresponds to the azimuthal direction with zero along the excitation axis of the antenna.
    
\begin{equation*}
            J_{\text{beam}, p}(\hat s,\nu) = \begin{bmatrix} 
        E_\theta^{px}(\hat s,\nu) & E_\phi ^{px} (\hat s,\nu)\\
        E_\theta ^{py}(\hat s,\nu) & E_\phi ^{py}(\hat s,\nu) 
        \end{bmatrix} ,
\end{equation*}
where $J_{beam}$ is a function of pointing direction $\hat s$ and frequency $\nu$. As seen above, the beam Jones matrix is calculated using the electric fields of both the orthogonal dipoles. It is important to note that the orthogonal dipole beams are placed as two rows in the matrix as given in \autoref{kroncker}. 
    
For the offset phase of the array, in addition to $J_{beam}$, we define another Jones matrix to capture the spatial offset between the phase centres. The offset between the orthogonal dipoles presents itself as an additional phase term that can be represented in the form of a Jones matrix:
    \begin{equation*}
 \begin{bmatrix} 
        \exp{(-i \psi)} & 0 \\
        0& \exp{(-i( 
    \psi + \Delta \psi))} \end{bmatrix} = \exp{(-i \psi)} \begin{bmatrix} 
        1 & 0 \\
        0& \exp{(-i
 \Delta \psi)} \end{bmatrix},
    \end{equation*}
where $-i \psi = -2\pi i (u l + v m)$ is a function of ($\hat s, \nu$) is taken into account by the $K_{pq}$ in Equation.4 and 6. So the Jones matrix due to the offset is given by:
\begin{equation}
    J_\text{offset} =  \begin{bmatrix} 
        1 & 0 \\
        0& \exp{(-i
 \Delta \psi)} \end{bmatrix},
 \label{eqn:jones-offset}
\end{equation}
where $\Delta \psi = 2\pi (\Delta u l + \Delta v m)$ captures the phase delay due to the Y-dipole being offset from the X-dipole.  In this case, the total Jones matrix is now given as: 
    \begin{equation}
          J = J_{\text{offset}} \times J_{\text{beam}}~.
          \label{total_J}
    \end{equation}

Next, we calculate the Muller matrices using the Jones matrix(ces) and the coordinate transform matrix S (Eq. 6):
    
    \begin{equation}
     M_{pq} = S^{-1}(J_p \otimes J_q^*)S ~,
     \label{muller-beam}
    \end{equation}
where we insert the appropriate J for the two cases: for the no-offset case, it would be just the $J_{Beam}$; for the offset phase case, the total J would be given by \autoref{total_J}.

Finally, we use Mueller matrix elements to calculate the fraction of Stokes sky ($\mathcal{E}^S = [I,Q,U,V]$) captured by the instrumental Pseudo Stokes parameters
    \begin{equation}
    \begin{split}
           {\mathcal{V}_{I}} = M00_{pq}I + M01_{pq}Q + M02_{pq}U + M03_{pq}V ,\\
          {\mathcal{V}_Q} =  M10_{pq}I + M11_{pq}Q + M12_{pq}U + M13_{pq}V ,\\
          {\mathcal{V}_{U}} = M20_{pq}I + M21_{pq}Q + M22_{pq}U + M23_{pq}V ,\\
            {\mathcal{V}_{V}} =  M30_{pq}I + M31_{pq}Q + M32_{pq}U + M33_{pq}V~.
               \end{split}
           \label{eqn:muller_matrix_intro}
           \end{equation}



\section{Simulating the Stokes Leakages for FARSIDE}
\label{sec:stokes_leakage}
\begin{figure*}
\centering
\includegraphics[width=\textwidth]{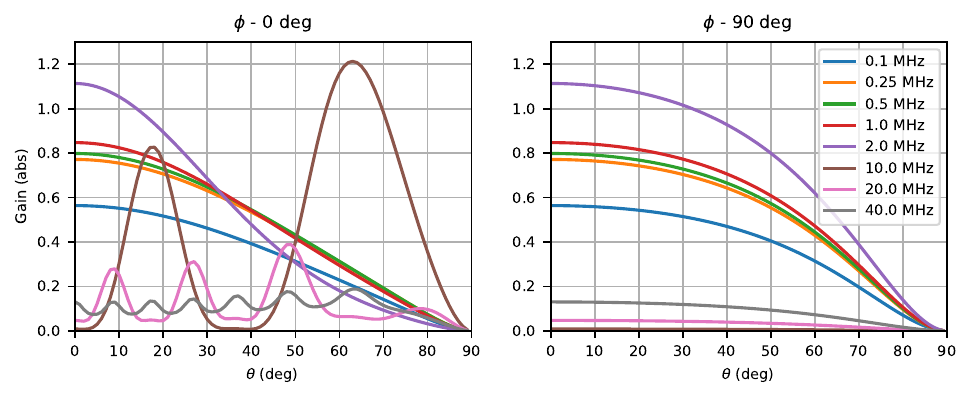}
\vspace*{-5ex}
\caption[Gain Curves of Dipoles on Regolith]{Simulated gain plots of a pair of co-located orthogonal 100~m dipole on regolith. Shown here is the gain vs. theta for one of the dipoles for a few frequencies in the FARSIDE operating band. The gains are shown at two cuts of azimuth ($\phi =0$ deg; along the excitation axis and $\phi=90$ deg, perpendicular to the excitation axis). Below 10~MHz, the beam patterns are donut-shaped with peak gain at the zenith. At higher frequencies, the beam pattern deviates from the ideal dipole-like pattern and is seen to have a multi-lobed response. }
\label{fig:dipole-beam-freqs}
\end{figure*}

We investigate the polarisation leakages caused by the antenna beams and offsets between the dipoles by simulating and calculating the Muller matrices explained above. To carry out this study, we obtain close-to-reality antenna beam models by simulating the orthogonal dipoles in a single antenna node with its exact dimensions (half-length = 50~m and radius = 1~mm) over regolith. This is done using an electromagnetic modelling software FEKO\footnote{Altair FEKO - https://altairhyperworks.com/product/FEKO}. We simulate two cases: one with the phase centres co-located and the other with the phase centres separated by 50~m. The regolith was modelled in both simulations at $Z=0$ using Green's function approximation with infinite extents in the $\pm$X, $\pm$Y, and -Z directions. We set the dielectric properties of the regolith using values of the lunar soil samples at $<1~$MHz from the Lunar Sourcebook, which reported the relative permittivity, $\epsilon_r = $ 2 and loss tangent, tan$\delta=10^{-3}$ \citep{Heiken1991}. We obtain the orthogonal E-field patterns (E$_{\theta}$ and E$_{\phi}$) and total power beams for both the X and Y dipoles. 
The gain cross-sections at $\phi=0^\circ$ and $\phi = 90^\circ$ of the power beam patterns for each simulated polarisation are shown in \autoref{fig:dipole-beam-freqs}. The gain versus zenith angle behaves as expected over the simulated frequencies for a 100m half-wavelength dipole. Close to the resonant frequency, at 2~MHz, the dipole has the maximum gain in both the gain planes. At frequencies $< 2~$MHz, the dipole is electrically short, so the gain reduces, but the overall beam pattern is larger, as seen in the similar half-beam widths in both the cross sections. At frequencies $>2~$ MHz, the dipole is electrically long, hence it starts showing a bifurcated beam response with increasing side lobes. This also increases the chromaticity of the beam (variation of the beam with frequency) and can make it difficult to account for the hydrogen cosmology science case. 

\subsection{Co-located phase centres}
The beam was simulated over a range of frequencies between 100 kHz and 40 MHz, and to quantify the polarisation leakages in detail, we selected one frequency in the band, 2~MHz, which is roughly at the multiplicative centre of the band and is also close to the resonant frequency of the dipole. With the generated electric field patterns, we now estimate the Stokes leakages by defining the Jones matrix, as shown in Step 1 of the methods section, and then using that to calculate the Muller matrix, as shown in Step 2. As a reference, the initial calculation is done for the case of co-located dipoles in the array, i.e., only the effect of the beam is considered. The Muller matrix evaluated using \autoref{muller-beam} in Step2 will result in:

\begin{equation}
    Jp \otimes Jq^* = \begin{bmatrix}
        E_{\theta}^{px} \overline{E_{\theta}^{qx}} & E_{\theta}^{px} \overline{E_{\phi}^{qx}} & E_{\phi}^{px} \overline{E_{\theta}^{qx}} & E_{\phi}^{px} \overline{E_{\phi}^{qx}} \\
        E_{\theta}^{px} \overline{E_{\theta}^{qy}} & E_{\theta}^{px} \overline{E_{\phi}^{qy}} &E_{\phi}^{px} \overline{E_{\theta}^{qy}} & E_{\phi}^{px} \overline{E_{\phi}^{qy}} \\
        E_{\theta}^{py} \overline{E_{\theta}^{qx}} & E_{\theta}^{py} \overline{E_{\phi}^{qx}}& E_{\phi}^{py} \overline{E_{\theta}^{qx}} & E_{\phi}^{py} \overline{E_{\phi}^{qx}} \\
        E_{\theta}^{py} \overline{E_{\theta}^{qy}} & E_{\theta}^{py} \overline{E_{\phi}^{qy}} & E_{\phi}^{py} \overline{E_{\theta}^{qy}} & E_{\phi}^{py} \overline{E_{\phi}^{qy}} 
    \end{bmatrix} .
\end{equation}

We multiply this equation by the inverse of the unitary matrix, $S^{-1}$:

\begin{strip}
    \rule[-2.5\baselineskip]{\dimexpr(0.5\textwidth-0.5
    \columnsep-0.6pt)}{0.6pt}

\begin{equation}
    S^{-1} (J_p \otimes J_q^*) = \frac{1}{2}\begin{bmatrix}
        E_{\theta}^{px} \overline{E_{\theta}^{qx}} +
        E_{\theta}^{py} \overline{E_{\theta}^{qy}} 
        & E_{\theta}^{px} \overline{E_{\phi}^{qx}} +
        E_{\theta}^{py} \overline{E_{\phi}^{qy}} 
        & E_{\phi}^{px} \overline{E_{\theta}^{qx}} +
        E_{\phi}^{py} \overline{E_{\theta}^{qy}} 
        & E_{\phi}^{px} \overline{E_{\phi}^{qx}} + 
        E_{\theta}^{px} \overline{E_{\theta}^{qx}}\\
        E_{\theta}^{px} \overline{E_{\theta}^{qx}} 
        -E_{\theta}^{py} \overline{E_{\theta}^{qy}} 
        & E_{\theta}^{px} \overline{E_{\phi}^{qx}} -
        E_{\theta}^{py} \overline{E_{\phi}^{qy}} 
        & E_{\phi}^{px} \overline{E_{\theta}^{qx}} -
        E_{\phi}^{py} \overline{E_{\theta}^{qy}} 
        & E_{\phi}^{px} \overline{E_{\phi}^{qx}} -
        E_{\phi}^{py} \overline{E_{\phi}^{qy}} \\
        E_{\theta}^{px} \overline{E_{\theta}^{qy}} + 
        E_{\theta}^{py} \overline{E_{\theta}^{qx}} 
        & E_{\theta}^{px} \overline{E_{\phi}^{qy}} + 
        E_{\theta}^{py} \overline{E_{\phi}^{qx}} 
        &E_{\phi}^{px} \overline{E_{\theta}^{qy}} + 
        E_{\phi}^{py} \overline{E_{\theta}^{qx}}
        & E_{\phi}^{px} \overline{E_{\phi}^{qy}} + 
        E_{\phi}^{py} \overline{E_{\phi}^{qx}} \\
        -i E_{\theta}^{px} \overline{E_{\theta}^{qy}} + 
        i E_{\theta}^{py} \overline{E_{\theta}^{qx}} 
        & -i E_{\theta}^{px} \overline{E_{\phi}^{qy}} + 
        i E_{\theta}^{py} \overline{E_{\phi}^{qx}} 
        & -i E_{\phi}^{px} \overline{E_{\theta}^{qy}} + 
        i E_{\phi}^{py} \overline{E_{\theta}^{qx}}
        & -i E_{\phi}^{px} \overline{E_{\phi}^{qy}} + 
        i E_{\phi}^{py} \overline{E_{\phi}^{qx}} 
        \end{bmatrix}.
\end{equation}


This is then multiplied by the unitary matrix S, so the total $S^{-1} (J_p \otimes J_q^*) S$ is:

\par
\begin{equation}
\resizebox{\textwidth}{0.07\textwidth}{$
M_{pq} = \frac{1}{2}\begin{bmatrix}
        E_{\theta}^{px} \overline{E_{\theta}^{qx}} + E_{\theta}^{py} \overline{E_{\theta}^{qy}} + E_{\phi}^{px} \overline{E_{\phi}^{qx}}  + E_{\phi}^{py} \overline{E_{\phi}^{qy}}
        &E_{\theta}^{px} \overline{E_{\theta}^{qx}} + E_{\theta}^{py} \overline{E_{\theta}^{qy}} - E_{\phi}^{px} \overline{E_{\phi}^{qx}}  - E_{\phi}^{py} \overline{E_{\phi}^{qy}}
        & E_{\theta}^{px} \overline{E_{\phi}^{qx}} + E_{\theta}^{py} \overline{E_{\phi}^{qy}}+ E_{\phi}^{px} \overline{E_{\theta}^{qx}} + E_{\phi}^{py} \overline{E_{\theta}^{qy}}
          & i E_{\theta}^{px} \overline{E_{\phi}^{qx}} + i E_{\theta}^{py} \overline{E_{\phi}^{qy}}-i E_{\phi}^{px} \overline{E_{\theta}^{qx}} -i E_{\phi}^{py} \overline{E_{\theta}^{qy}} \\
          E_{\theta}^{px} \overline{E_{\theta}^{qx}} - E_{\theta}^{py} \overline{E_{\theta}^{qy}} +E_{\phi}^{px} \overline{E_{\phi}^{qx}}  - E_{\phi}^{py} \overline{E_{\phi}^{qy}}
         &  E_{\theta}^{px} \overline{E_{\theta}^{qx}} - E_{\theta}^{py} \overline{E_{\theta}^{qy}} -E_{\phi}^{px} \overline{E_{\phi}^{qx}}  + E_{\phi}^{py} \overline{E_{\phi}^{qy}}
        & E_{\theta}^{px} \overline{E_{\phi}^{qx}} - E_{\theta}^{py} \overline{E_{\phi}^{qy}}+
        E_{\phi}^{px} \overline{E_{\theta}^{qx}} - E_{\phi}^{py} \overline{E_{\theta}^{qy}}
        & i E_{\theta}^{px} \overline{E_{\phi}^{qx}} - i E_{\theta}^{py} \overline{E_{\phi}^{qy}}
        -i E_{\phi}^{px} \overline{E_{\theta}^{qx}} + i E_{\phi}^{py} \overline{E_{\theta}^{qy}}\\
        E_{\theta}^{px} \overline{E_{\theta}^{qy}} + E_{\theta}^{py} \overline{E_{\theta}^{qx}}+ E_{\phi}^{px} \overline{E_{\phi}^{qy}} + E_{\phi}^{py} \overline{E_{\phi}^{qx}} 
        &E_{\theta}^{px} \overline{E_{\theta}^{qy}} + E_{\theta}^{py} \overline{E_{\theta}^{qx}} -E_{\phi}^{px} \overline{E_{\phi}^{qy}} - E_{\phi}^{py} \overline{E_{\phi}^{qx}}
        & E_{\theta}^{px} \overline{E_{\phi}^{qy}} + E_{\theta}^{py} \overline{E_{\phi}^{qx}}+
        E_{\phi}^{px} \overline{E_{\theta}^{qy}} + E_{\phi}^{py} \overline{E_{\theta}^{qx}}
        & i E_{\theta}^{px} \overline{E_{\phi}^{qy}} + i E_{\theta}^{py} \overline{E_{\phi}^{qx}}
        -i E_{\phi}^{px} \overline{E_{\theta}^{qy}} - i E_{\phi}^{py} \overline{E_{\theta}^{qx}}\\
        -i E_{\theta}^{px} \overline{E_{\theta}^{qy}} + iE_{\theta}^{py} \overline{E_{\theta}^{qx}}
        -i E_{\phi}^{px} \overline{E_{\phi}^{qy}} +i E_{\phi}^{py} \overline{E_{\phi}^{qx}}
       &-i E_{\theta}^{px} \overline{E_{\theta}^{qy}} +i E_{\theta}^{py} \overline{E_{\theta}^{qx}}
        +i E_{\phi}^{px} \overline{E_{\phi}^{qy}} -i E_{\phi}^{py} \overline{E_{\phi}^{qx}}
        & -i E_{\theta}^{px} \overline{E_{\phi}^{qy}} + i E_{\theta}^{py} \overline{E_{\phi}^{qx}}
        -i E_{\phi}^{px} \overline{E_{\theta}^{qy}} +i E_{\phi}^{py} \overline{E_{\theta}^{qx}}
        &  E_{\theta}^{px} \overline{E_{\phi}^{qy}} -  E_{\theta}^{py} \overline{E_{\phi}^{qx}}
        - E_{\phi}^{px} \overline{E_{\theta}^{qy}} + E_{\phi}^{py} \overline{E_{\theta}^{qx}} \end{bmatrix}$}~.
\end{equation}.
\hfill
\rule[-0.6\baselineskip]{\dimexpr(0.5\textwidth-1\columnsep-0.6pt)}{0.6pt}%
\end{strip}

The absolute values of the M$_{pq}$ at 2~MHz for the co-located case are plotted in \autoref{fig:stokes-beam}a to estimate the fractional leakages between the various Stokes components. The Muller matrices are projected in the $\theta/\phi$ basis, and all of the dynamic ranges are normalised to the peak of M$_{00}$, which is one at the zenith. The off-diagonal Muller matrix components would be zero for an ideal instrument with no polarisation leakage/mixing. The first column corresponds to the sky Stokes I coupling into the instrument's all four Stokes components (I $\rightarrow \mathcal{V}^I, \mathcal{V}^Q,\mathcal{V}^U,\mathcal{V}^V$ ). Similarly, the second, third, and fourth columns capture the sky's Stokes Q, U, and V components into all four pseudo-Stokes visibilities of the instrument. At low frequencies and large scales probed by many low-frequency interferometers, Stokes I of the sky is extremely bright compared to the other Stokes parameters. In early studies, only a few polarised point sources have been observed at frequencies below 300~MHz \citep{Bernardi2013, asad2016} and low-level diffused polarised emission, which peaks at 4K at 154 MHz \citep{Lenc_2016,Bryne_2022}. However, more recent papers by \cite{sullivan_2023, Eck_2018} analysing the LOTS-DR2 LOFAR data (120-168~MHz) have detected more polarised point sources. But they also show that these are fewer in number compared to the NVSS survey (1.4~GHz) in a given sky area. In addition, \cite{sullivan_2023} find that the median degree of polarisation (1.8\%) is about 3 to 10 times lower than those estimated with the NVSS survey. Applying a similar depolarisation scaling to FARSIDE frequencies, we can expect the degree of polarisation to be < 1\%. This is similar to the findings in \cite{Farnes2014}, which showed evidence for systematic depolarisation of steep-spectrum point sources towards low frequencies, causing low polarisation fractions ($\leq$1\%) below 300~MHz. At FARSIDE frequencies, we would be primarily concerned about sky Stokes I leaking into other instrumental Stokes components. The Muller matrices indicate 20\%, 4\% and 0.075\% fractional leakages of Stokes I into $\mathcal{V}_Q$, $\mathcal{V}_U$, and, $\mathcal{V}_V$ respectively, for the co-located case. 

For the exoplanet science case, we are concerned about the Muller matrices M$_{30}$, M$_{31}$ and M$_{32}$; leakage of the sky components into the measured pseudo-Stokes $\mathcal{V}_V$ (bottom row of \autoref{fig:stokes-beam}). In this case, with no spatial offset and only direction-dependent beam effects, we find the leakage to be at least two~orders magnitude lower compared to M$_{33}$ which is the instrumental pseudo-Stokes $\mathcal{V}_{V}$ capturing the sky Stokes V. For the hydrogen cosmology science case, we are interested in the sky Stokes components Q, U and V leaking into the $\mathcal{V}_I$. This is quantified by M$_{01}$, M$_{02}$ and M$_{03}$ and maximum percentages from each are 20\%,7.5\% and 0.002\% respectively. So, the major contribution is from the Stokes Q leaking into the instrumental Stokes I. Due to the direction-dependent beam of the dipoles on the lunar regolith, we see considerable leakage of the polarised sky into the unpolarised power received. 


\subsection{Non-colocated phase centres}
To study the polarisation leakages due to spatial offsets between the orthogonal dipole pairs of an antenna, we follow the same procedure as above and define a Jones matrix to account for the differential delay between the signal received by the X and Y dipoles (as shown in \autoref{muller_steps}). The Muller matrix accounting for only this offset (excluding the antenna beam contribution) evaluates to:

\begin{equation}
    J \otimes J^* = \begin{bmatrix}
        1 & 0 & 0 & 0  \\
        0 & \exp{(i \Delta \psi)} & 0 & 0 \\
        0 & 0 & \exp{(-i \Delta \psi)} & 0 \\
        0 & 0 & 0 & 1
    \end{bmatrix},
\end{equation}

\begin{equation}
    S^{-1}(J \otimes J^*) = \frac{1}{2}\begin{bmatrix}
        1 & 0 & 0 & 1  \\
        1 & 0 & 0 & -1 \\
        0 &  \exp{(i \Delta \psi)} & \exp{(-i \Delta \psi)} & 0 \\
        0 &  -i \exp{(i \Delta \psi)} & i \exp{(-i \Delta \psi)} & 0
    \end{bmatrix},
\end{equation}

\begin{equation}
    M_\text{offset} = \begin{bmatrix}
        1 & 0 & 0 & 0  \\
        0 & 1 & 0 & 0 \\
        0 &  0&  \cos{\Delta \psi} &  -\sin{\Delta \psi} \\
        0 &  0& \sin{\Delta \psi} &   \cos{\Delta \psi}
    \end{bmatrix}~.
    \label{eqn:only offset}
\end{equation}

\begin{figure*}
\begin{subfigure}{0.49\textwidth}
\includegraphics[width=0.90\textwidth]{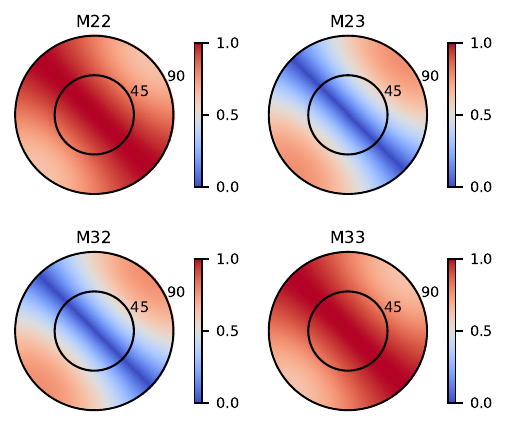}
\caption{0.6~MHz}
\end{subfigure}
\begin{subfigure}{0.49\textwidth}
\includegraphics[width=0.90\textwidth]{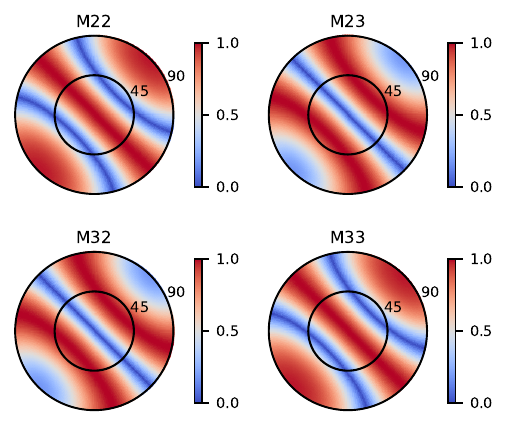}
\caption{2~MHz}
\end{subfigure}

\centering
\begin{subfigure}{0.49\textwidth}
\includegraphics[width=0.90\textwidth]{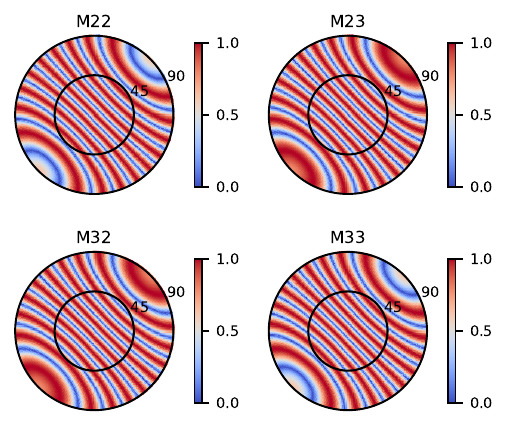}
\caption{10~MHz}
\end{subfigure}
\caption[Muller Matrix coefficients for Offset Case]{Muller matrices that are non-zero when only the non-colocated phase centre effect is considered, excluding the beam pattern contribution. These indicate that direction-dependent intermixing of only Stokes U and V sky components occurs due to the spatial offset between the phase centres of the orthogonal dipoles. The effects are shown for three frequencies: 0.6~MHz, 2~MHz, and 10~MHz.}
\label{fig:only_offset}
\end{figure*}

The case of only offset is seen to affect $ M_{22}, M_{23}, M_{32}~\rm{and}~M_{33} $, i.e., the coefficients that quantify the leakage from the sky Stokes U and V to the instrument's Stokes $\mathcal{V}_U$ and $\mathcal{V}_V$. This is seen in \autoref{eqn:only offset} and \autoref{fig:only_offset}. In \autoref{fig:only_offset}, we plot the relevant Muller matrix coefficients for a few frequencies to assess how the offset effects change with frequency. Confirming what was obtained in the M$_{\rm{offset}}$ equation, the $M_{22}$ and $M_{33}$ are the same and the $M_{23}$ and $M_{32}$ are the same in magnitude. As the frequency increases, the delay due to the offset, which depends on the ratio of the offset value to the observation wavelength, undergoes multiple cycles. In the case of 0.6~MHz (or 500~m wavelength), the offset is only 0.1$\lambda$ in the u and v directions, and the exact number of cycles of variations between 0 and 1 can be understood by the total delay given as:
\begin{equation*}
   cos\{\Delta \psi\} = cos \{2 \pi \frac{50}{\lambda} (sin \theta (cos \phi + sin \phi))\}.
\end{equation*}
To obtain the number of cycles, we calculate the number of times $cos\{\Delta \psi\} $ evaluates to 1, i.e., what is the integer multiple of 2$\pi$ possible in the above equation for different wavelengths/frequencies. To do so, we need the $\theta$ and $\phi$ at which the maximum value is possible, and substituting $\theta = 90^\circ$ and $\phi = 45^\circ$ gives us that.
\begin{equation*}
    cos\{\Delta \psi\}_{max} = cos\{2 \pi\frac{50}{\lambda} \sqrt{2}\}~.
\end{equation*}

This implies it will have approximately 1/4 cycle for 0.6~MHz, one cycle for 2~MHz, and five cycles for 10~MHz.

\begin{figure*}
\centering
\includegraphics[width=0.9\textwidth]{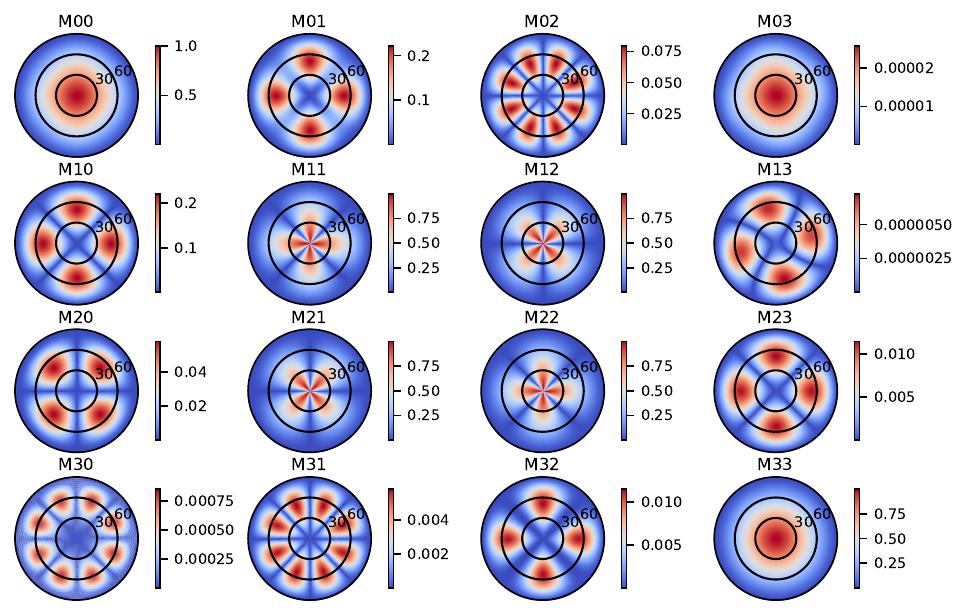}
\hrule
\centering
\includegraphics[width=0.9\textwidth]{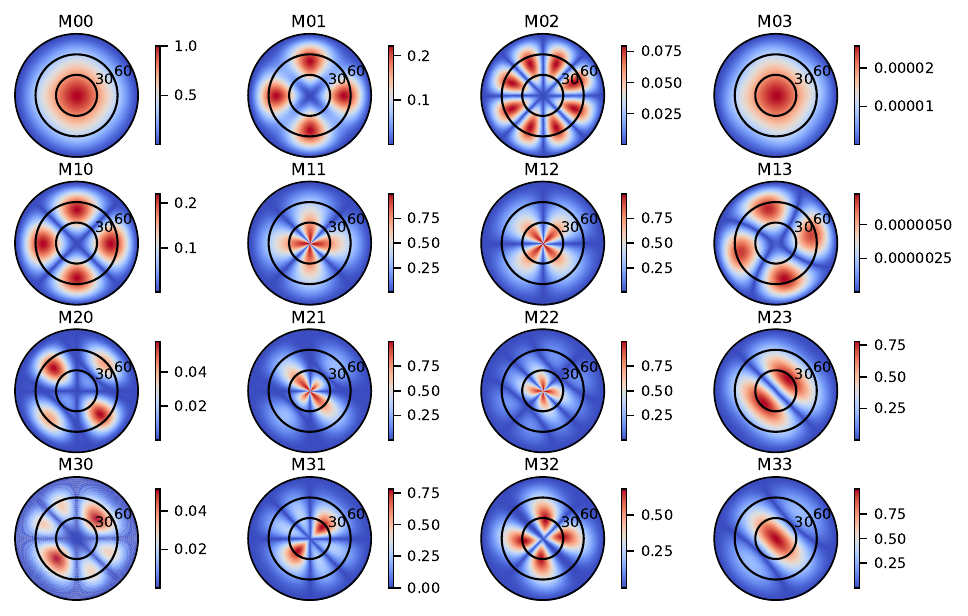}
\vspace*{-2.5ex}
\caption[Muller Matrix Plots of Colocated and offset phase Orthogonal Dipoles]{Plots of Muller matrix elements for simulated co-located [top] and non-colocated [bottom] cross dipoles on regolith at 2~MHz as a function of elevation angle ($\theta$) and azimuth angle ($\phi$). The absolute values of the elements are plotted. Colour bar scales are relative to the peak of M$_{00}$ (normalised to 1 at the zenith). The elements capture the fractional leakages of the sky Stokes components [I, Q, U, V] into the instrumental Psuedo-Stokes [$\mathcal{V}_I$,$\mathcal{V}_Q$, $\mathcal{V}_U$, $\mathcal{V}_V$]. For example, the first column corresponds to the sky Stokes I coupling into the instrument's all four Stokes components (I $\rightarrow \mathcal{V}^I, \mathcal{V}^Q,\mathcal{V}^U,\mathcal{V}^V$ ). See equation \ref{eqn:muller_matrix_intro} for a key to these matrices.}
\label{fig:stokes-beam}
\end{figure*}

Next, we combine the effects of offset and the direction-dependent beam so the joint Jones matrix is now defined as:
\begin{equation}
    J_\text{offset} \times J_\text{beam} = \begin{bmatrix}
        E_{\theta}^x & E_{\phi}^x \\
        \exp{(-i \Delta \psi)}E_{\theta}^y & \exp{(-i \Delta \psi)} E_{\phi}^y
    \end{bmatrix}~.
\end{equation}
Then the full $J \otimes J^*$ is calculated as above: 

\begin{strip}
    \rule[-2\baselineskip]{\dimexpr(0.5\textwidth-1\columnsep-0.6pt)}{0.6pt}%
\par
\begin{equation}
    ( J_\text{offset} \times J_\text{beam})  \otimes   (J_\text{offset} \times J_\text{beam})^*  = 
     \begin{bmatrix}
        E_{\theta}^x \overline{E_{\theta}^x} 
        & E_{\theta}^x \overline{E_{\phi}^x} 
        & E_{\phi}^x \overline{E_{\theta}^x} 
        & E_{\phi}^x \overline{E_{\phi}^x} \\
        \exp{(i \Delta \psi)} E_{\theta}^x \overline{E_{\theta}^y} 
        & \exp{(i \Delta \psi)} E_{\theta}^x \overline{E_{\phi}^y} 
        & \exp{(i \Delta \psi)} E_{\phi}^x  \overline{E_{\theta}^y} 
        & \exp{(i \Delta \psi)}E_{\phi}^x \overline{E_{\phi}^y} \\
        \exp{(-i \Delta \psi)}E_{\theta}^y \overline{E_{\theta}^x}
        & \exp{(-i \Delta \psi)}E_{\theta}^y \overline{E_{\phi}^x}
        & \exp{(-i \Delta \psi)}E_{\phi}^y \overline{E_{\theta}^x} 
        & \exp{(-i \Delta \psi)}E_{\phi}^y \overline{E_{\phi}^x} \\
        E_{\theta}^y \overline{E_{\theta}^y} 
        & E_{\theta}^y \overline{E_{\phi}^y}
        & E_{\phi}^y \overline{E_{\theta}^y}
        & E_{\phi}^y \overline{E_{\phi}^y} 
   \end{bmatrix}~.
    \label{eqn: joint-jones}
\end{equation}
\hfill
\rule[0.5\baselineskip]{\dimexpr(0.5\textwidth-0.5\columnsep-1pt)}{0.4pt}
\end{strip}

The joint Jones matrix of offset and beam is seen to have an additional delay factor multiplying the electric field patterns of the y-dipole (\autoref{eqn: joint-jones}). Thus, the resulting combined beam and offset Muller matrix affects every matrix element with an x- and a y-electric field component. Referring back to the Muller matrix (M$_{\rm{beam}}$), that would imply every element that determines the instrumental Psuedo-Stokes $\mathcal{V}_U$ and $\mathcal{V}_V$. This is seen in the last two rows of the bottom plot in  \autoref{fig:stokes-beam}, which shows the Muller matrices of the beam + offset at 2~MHz.

The offset does not add additional leakages or mixing in the collected Stokes I ($\mathcal{V}_I$); thus, it does not directly affect 21~cm measurements. But it causes more mixing of other sky Stokes components into the acquired Stokes~V ($\mathcal{V}_V$). This affects the exoplanet science case where we want to capture the pure Stokes V from the electron cyclotron maser emission of the star's type~II bursts and the host-planet interaction. Specifically the max values of M$_{30}$ increases from 0.07\% to 4\%, M$_{31}$ from 0.4\% to 75\% and  M$_{32}$ from 1\% to 50 \%. 

\subsection{Frequency dependence}

\begin{figure*}[h!]
    \centering
    \includegraphics[width=\textwidth]{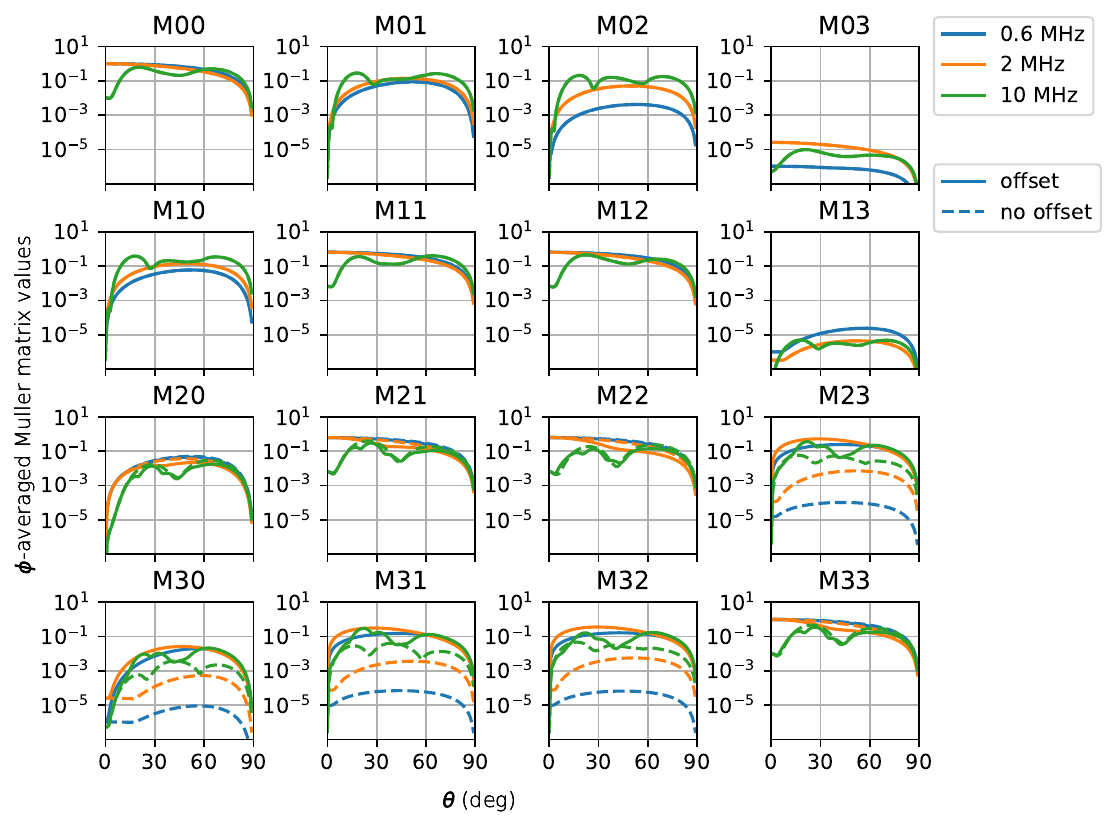}
    \caption[Stokes Leakage Varying with Frequency]{Azimuthally averaged absolute Muller matrix values for spatially co-located[dashed curves] and non-co-located [solid curves] orthogonal dipoles versus elevation angle. This is shown for three different frequencies within the operating band of the FARSIDE array. The effect of the offset is only on the last two rows of the Muller matrix. With increasing frequency, the absolute fractional mixing due to offset decreases. At the frequency of 10~MHz, the beam pattern of the antenna is no longer in the ideal dipole regime.}
    \label{fig:stokes_freq}
\end{figure*}

The pseudo-Stokes leakages due to the beam and offset analysed so far (\autoref{fig:stokes-beam}) were only for observations at 2~MHz. We know from \autoref{fig:dipole-beam-freqs} that the beam pattern varies with frequency, and the delay due to the offset also varies with frequency(\autoref{fig:only_offset}). Thus, we can expect the Stokes leakage to vary in frequency.  We calculated the Muller matrices similar to \autoref{fig:stokes-beam} for frequencies 0.6~MHz and 10~MHz, which approximately capture the low and high ends of the FARSIDE bandwidth. We then take the azimuthal averages of these absolute Muller matrices of each frequency and plot the result in \autoref{fig:stokes_freq}. The rows and columns correspond to respective Muller matrix positions (similar to \autoref{fig:stokes-beam}), with the rows corresponding to instrumental Stokes components($\mathcal{V}$) and the columns to the sky Stokes components. Each subplot is the Stokes leakages (normalised to a maximum value of $M_{00}$) averaged along the azimuth axis, and the individual curves correspond to different frequencies: 0.6, 2, and 10~MHz. Considering just the co-located dipoles case, the plot shows that with an increase in frequency, the Muller matrix values also increase. And at 10~MHz, we notice higher variations with theta(viewing angle) as expected from \autoref{fig:dipole-beam-freqs}. When including a phase offset, similar to before, the offset effects increase in the leakages in the instrumental pseudo-Stokes $\mathcal{V}_U$ and $\mathcal{V}_V$. But with an increase in frequency, the factor of increase in Stokes leakages is seen to decrease.    

\subsection{Intrinsic Cross-polarisation Metrics}
We quantify the polarimetry performance of FARSIDE in terms of its \textit{intrinsic cross-polarisation ratio} (IXR$_M$) for Stokes polarimetry in terms of Muller calculus defined by \cite{carozzi2011,kohn2019,asad2016}. This metric is independent of the choice of the coordinate systems, unlike the standard IEEE definition of cross-polarisation ratio. It is defined as:
\begin{equation}
    IXR_M = \frac{1}{D} = \frac{|M_{00}|}{\sqrt{M_{10}^2+M_{20}^2+M_{30}^2}}~.
\end{equation}
This expression quantifies the instrumental polarisation (D), i.e., leakage of the unpolarised power into polarised power due to the instrument. 

We adapt this metric and define two Figure of Merits that will appropriately capture the performance of FARSIDE for its two desired science cases:
\\\\1.) Quantifying leakage into $\mathcal{V}_I$
\begin{equation}
    IXR_{M,I} = \frac{\sqrt{M_{01}^2 + M_{02}^2 + M_{03}^2}}{|M_{00}|}~.
    \label{eqn:ixr_mi}
\end{equation}

This will quantify how much of the sky Stokes Q, U, and V gets depolarised and leaks into instrumental Stokes I ($\mathcal{V}_I$). This is important in the case of the Dark Ages science since the faint redshifted 21cm signal is expected to be unpolarised, and the brighter foreground will be partially polarised.
\\\\2.)Quantifying leakage into $\mathcal{V}_V$
\begin{equation}
    IXR_{M,V}  = \frac{\sqrt{M_{30}^2 + M_{31}^2 + M_{32}^2}}{|M_{33}|}~.
    \label{eqn:ixr_mv}
\end{equation}

This will quantify how much of the sky Stokes I, U and Q leaks into instrumental Stokes I ($\mathcal{V}_I$). This is important in the case of the exoplanet magnetosphere emission science since the electron cyclotron maser mechanism is expected to produce 100\% circular polarised radiation.

\begin{figure*}
    \centering
    \includegraphics[width=\textwidth]{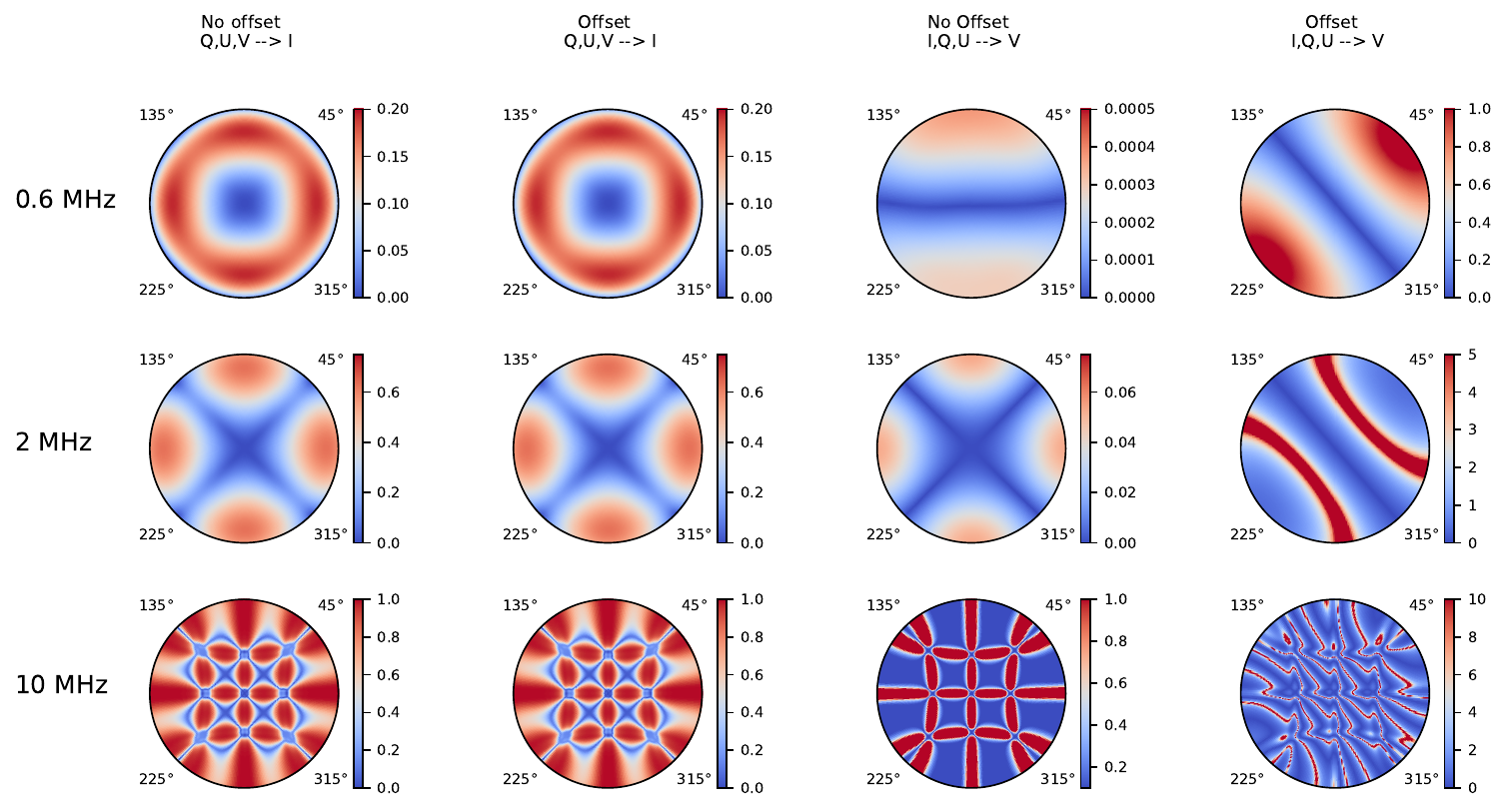}
    \caption[Intrinsic Cross Polarisation for the Offset and No Offset Cases]{Intrinsic cross polarisation values for Stokes I and V for cases of no offset and with offset. Each row corresponds to a different frequency. The offset doesn't add to the leakage into Stokes I, but it affects the leakages into Stokes V, which increases with frequency.}
    \label{fig:fom_freq}
\end{figure*}

 We plot these crosspolarisationn ratios in \autoref{fig:fom_freq} for three frequencies in the FARSIDE's planned bandwidth - 0.6~MHz, 2~MHz and 10~MHz. The IXM values increase with frequency. The IXM$_{M,I}$ is not affected by the phase offset between the dipoles. The offset increases IXM$_{M,V}$ by factors of 2000, 100, and 10 at 0.6~MHz, 2~MHz, and 10~MHz, respectively. The Stokes leakage factors due to the offset are seen to decrease in frequency, as was noted with the azimuthally averaged plots too. 

\section{Simulating Observations}
\label{sec:effects_onsky}
This section investigates the effects of FARSIDE direction-dependent beam and spatial phase offsets on simulated sky observations. Until the previous section, we quantified the leakage fractions due to the beam and offset of the dipoles via Muller matrices. In this section, we use the calculated Muller matrices and \autoref{eqn:vis-final} to simulate dirty(un-deconvolved) images of the sky recovered by FARSIDE. We note that these polarisation leakages can be compensated for with calibration if the beam variation and offsets are known accurately. Any errors in the offsets will lead to errors in the calibration, which will, in turn, result in polarisation leakages. In section \ref{sec:correction}, we present a simple correction that can be applied to deconvolved images to compensate for the offset effects if the dipole positions are known accurately. As shown in the equation, we use the sky Stokes matrix($\mathcal{E}^S = [I~Q~U~V]^T$) and the \textit{uv-}coverage of the array to process the model sky through the FARSIDE array and obtain the pseudo-Stokes images. We input only a Stokes~I sky model and look at the obtained fractional leakages in other Stokes components. Thus, the simulated observations will be limited, but we are interested in the V/I ratio, i.e., how much of the sky's Stokes~I leaks into the instrumental pseudo-Stokes~V image.

\begin{figure*}
    \centering
    \includegraphics[width=1\textwidth]{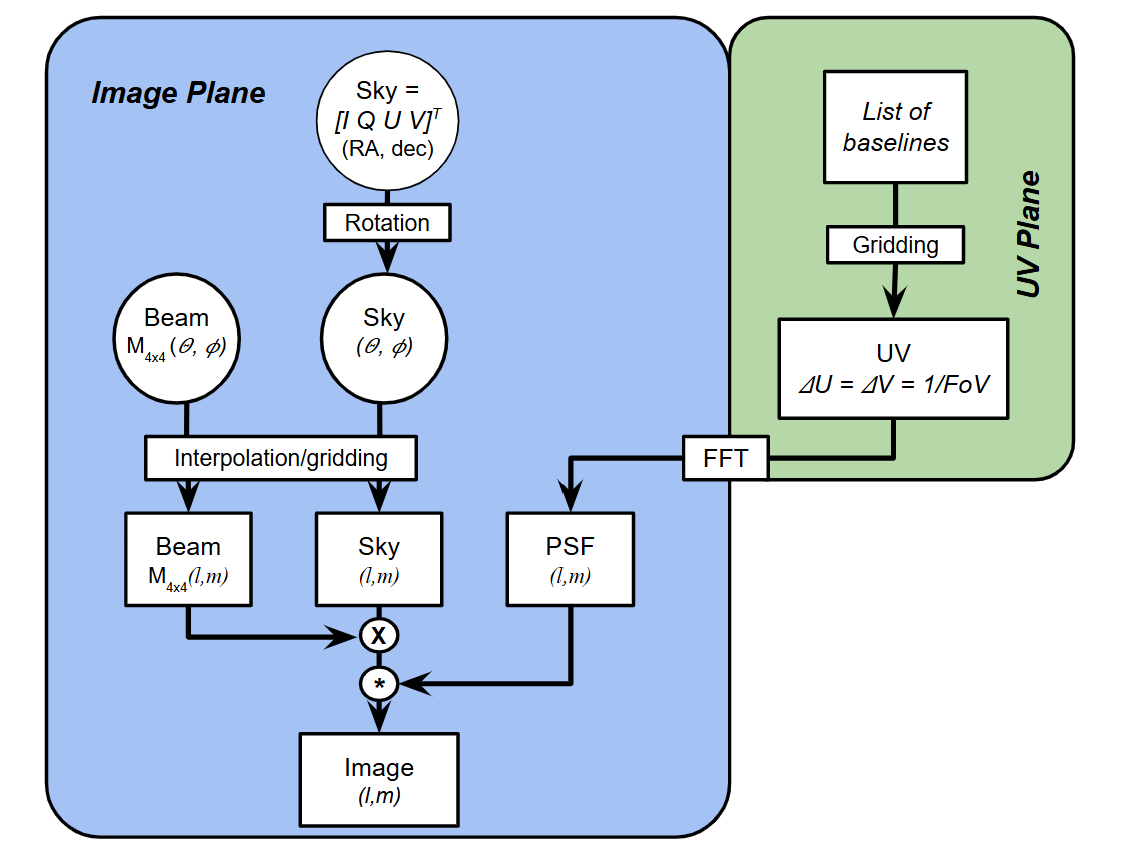}
    \caption[FARSIDE Imagining Pipeline Diagram]{Block diagram of the complete flow of the interferometer pipeline developed for FARSIDE. It shows how the sky images are pre-processed before multiplying the beam. For the beam, the flowchart indicates the Muller matrices that take into account the beam and the spatial offset effects of the dipoles of the array.}
    \label{fig:pipeline}
\end{figure*}

The complete process followed in the pipeline is shown in \autoref{fig:pipeline}. First, we process the sky. We take the sky model consisting of point sources and convert its positions from Right Ascension (RA) and declination (dec) to elevation($\theta$) and azimuth ($\phi$) with reference to the location of the array. The time chosen for the viewing is when the Galactic Centre is not in the field of view of the FARSIDE array. Both the two-dimensional Muller matrices and sky arrays are functions of $\theta$ and $\phi$ in spherical coordinates. We interpolate those to a uniform grid of $l = sin(\theta)cos(\phi)$ and  $m = sin(\theta)sin(\phi)$. The number of points per side on the l and m grid ($N_{l,m}$) is decided by the resolution of the array at the observed frequency by:
\begin{equation*}
\begin{split}
    N_{l,m} &= \frac{2 \pi}{resolution} \\
    & = 2 \pi d(\lambda)~.
\end{split}
\end{equation*}
where $d(\lambda)$ is the longest baseline in terms of wavelength. At 2~MHz, the longest baseline is $\approx$80 $\lambda$. Thus, we use 500 x 500 pixels for the images at 2 MHz ($N_{l,m} = 500$). We assign the interpolated sky array to the sky Stokes matrix. Suppose the sky has only Stokes I components, as in our case here, the $\mathcal{E}^S = [I(l,m)~0~0~0]^T$. We simplify the problem by limiting our field of view to 30$^\circ$ around zenith ($sin(\pi/6)\approx\pi/6$) and assuming no $w$-effect. Referencing \autoref{eqn:vis-final} again, we multiply the truncated and interpolated Muller matrices and sky Stokes matrix. To convolve with the point spread function (PSF) of the array, we Fourier transform this product and multiply it with the \textit{uv-}coverage. 

We calculate the \textit{uv-}coverage for a zenith-pointing XY array (at RA = 0 and dec = 0) at a location on the Earth's equator. This assumption has to be carefully translated to an array on the FARSIDE of the Moon. Given that the equator of the Moon is offset by a maximum of 28.7$^\circ$ and that the final location of the array is chosen to be within $\pm 30 ^\circ$ from the Moon's equator, our assumption of a zenith-pointing array at the Earth’s equator yields a reasonable example field for FARSIDE. In this analysis, we are only considering snapshot imaging.   We note that for rotation synthesis, the sky rises and sets at a much longer timescale on the Moon than on Earth (28 days vs 24 hours), but otherwise, it enables similar outcomes. The \textit{uv-}coverage simplifies to:

\begin{equation*}
\begin{split}
    u\approx cos(RA) * x = x,\\                
    v\approx cos(dec)*y + sin(dec)*sin(RA)*x =y .\\ 
\end{split}
\end{equation*}
where \textit{x} and \textit{y} are the XX and YY baselines, respectively.

The $uv-$coverage, given 128 antennas, has $128\times127=16256$ \text{uv-}points and is binned into a two-dimensional histogram with the number of bins same as the number of pixels in the \textit{l,m} grid (500 pixels). The binned \textit{uv-}coverage is multiplied by the Fourier-transformed product and inverted back to the image domain. This yields dirty images with natural-weighted PSFs.
This process is repeated for both the co-located and offset phase cases.

\bigskip
\subsection{Sky Model}

\begin{figure*}
\centering
\includegraphics[width=0.5\textwidth]{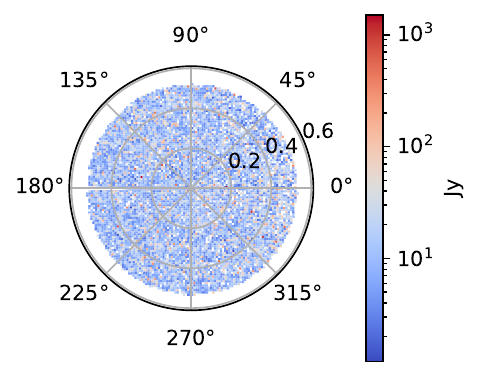}
\caption[GLEAM Point Source Model]{The stokes I sky model using the GLEAM point sources between 1Jy and 1.5kJy within 30 deg of the zenith over the assumed FARSIDE phase center (ra = dec =0$^\circ$). These sources are projected onto the same \textit{l} and \textit{m} grid as the beams of the FARSIDE array. This Stokes I sky model is processed following \autoref{sec:effects_onsky} to yield simulated observations (see \autoref{fig:GLEAM_image})}
\label{fig:GLEAM_sky_model}
\end{figure*}

\begin{figure*}
\centering
\includegraphics[width=\textwidth]{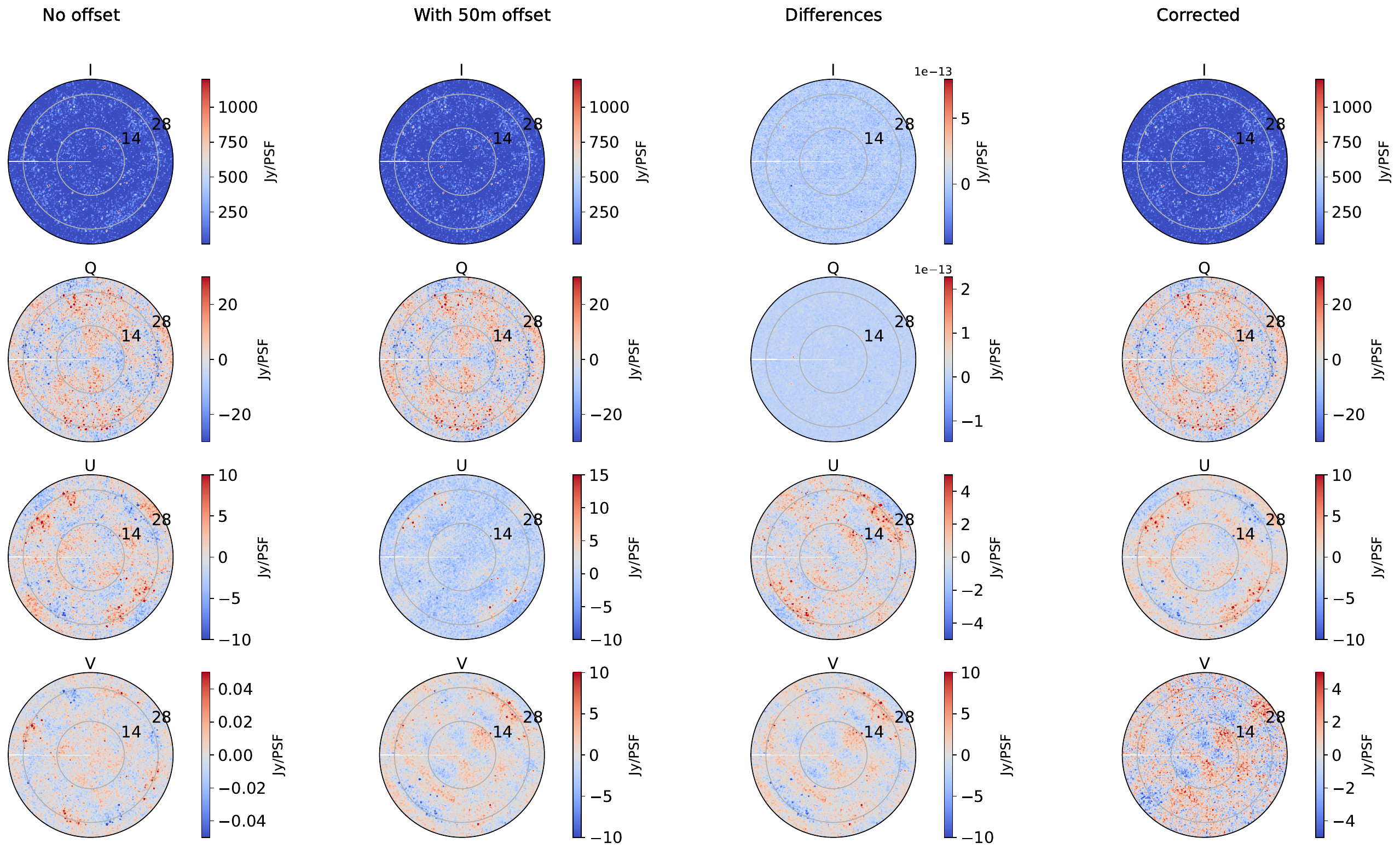}
\vspace*{-5ex}
\caption[Simulated Images of GLEAM through the Pipeline]{Simulated I, Q, U, and V images of the sources from the GLEAM catalogue through the developed pipeline that includes the beam, the snapshot PSF, and phase offset. The images are generated at the centre of the FARSIDE band, i.e., 2~MHz. The five image columns correspond to: co-located X and Y dipoles, with 50m offset between the X and Y dipoles, differences between co-located and non-colocated cases and offset corrected image.}
\label{fig:GLEAM_image}
\end{figure*}

For the model sky, we use the GLEAM point source catalogue\footnote{https://heasarc.gsfc.nasa.gov/W3Browse/all/gleamegcat.html} surveyed by the Murchinson Widefield Array \citep{hurleywalker2017}. We note that this model does not include any extended/diffuse emission, and we will leave the analysis of that to future work. The GLEAM survey has flux values for radio sources between 72-230~MHz. We read in all the point sources and their associated spectral indices from the GLEAM catalogue using the pyradio sky python package\footnote{https://github.com/RadioAstronomySoftwareGroup/pyradiosky}. Using pyradiosky, the spectral index, and the radio flux at 72~MHz (lowest frequency of data available with GLEAM), we estimate the radio flux at 2~MHz of each source and limit the number of sources by restricting the flux range between 1~Jy and 1.5~kJy. This sky that is clipped to 30$^\circ$ from zenith is shown in \autoref{fig:GLEAM_sky_model}. We feed in only the Stokes I components of this output as the model sky into the pipeline described above. We obtain all four instrumental pseudo-Stokes dirty images for the FARSIDE array layout's colocated and offset phase dipoles. The resulting images for both cases are shown as the first two columns of \autoref{fig:GLEAM_image}.

As expected, the pseudo-Stokes I and Q images are identical for the colocated and spatially phase offset cases. The offset affects only the acquired stokes U and V images. The direction-dependent beam results in maximum V/I to be $\approx$ 5$*10^{-5}$. Adding a spatial offset to the phase centres of the dipoles with this direction-dependent beam increases the maximum V/I to $\approx$8$*10^{-3}$, i.e., a two orders of magnitude increase in the leakage. The third column shows the differences between the colocated and non-colocated cases for each Stokes image. The differences are more prominent for viewing angles $>10^\circ$ from the zenith. 

\subsection{Correcting for the Offset}
\label{sec:correction}

Preliminary analysis of modelling a pure Stokes I sky through the FARSIDE beam and array shows that the offset between the phase centres of the dipoles within each node leads to considerable leakage into the instrumental pseudo-Stokes $\mathcal{V}_U$ and $\mathcal{V}_V$ dirty images. The increased leakage into Stokes $\mathcal{V}_V$ increases the confusion noise in searches for the circularly polarised emission from magnetospheres of exoplanets. Therefore, it is crucial to account for the offset's effects. Here, we present a formulation for the correction. 

The constructed pseudo-Stokes visibilities (ignoring the PSF of the array) from the model sky for co-located case (no spatial offset) are given by:

\begin{equation*}
    \begin{bmatrix}
        \mathcal{V}_I \\
        \mathcal{V}_Q \\
        \mathcal{V}_U \\
        \mathcal{V}_V 
    \end{bmatrix} = M_{\rm{beam}} \begin{bmatrix}
         I \\
        Q \\
        U \\
        V 
    \end{bmatrix}
\end{equation*}
\begin{equation*}
   = S^{-1} (J_{\rm{beam}})\otimes(J_{\rm{beam}})^* S \begin{bmatrix}
         I \\
        Q \\
        U \\
        V
    \end{bmatrix}~.
\end{equation*}

Applying some math to enable the offset correction. This can be rewritten as: 
\begin{equation*}
\begin{split}
   \begin{bmatrix}
        \mathcal{V}_I \\
        \mathcal{V}_Q \\
        \mathcal{V}_U \\
        \mathcal{V}_V 
    \end{bmatrix} = &S^{-1} (J_{\rm{offset}}^{-1}\times J_{\rm{offset}}\times J_{\rm{beam}})\\
    &\otimes(J_{\rm{offset}}^{-1}\times J_{\rm{offset}}\times J_{\rm{beam}})^* S \begin{bmatrix}
         I \\
        Q \\
        U \\
        V 
    \end{bmatrix}~.
\end{split}
\end{equation*}

Applying the Kronecker product rule again;
\begin{equation*}
\begin{split}
   =& S^{-1} (J_{\rm{offset}}^{-1}\otimes (J_{\rm{offset}}^{-1})^*) \\
   &\times (J_{\rm{offset}}\times J_{\rm{beam}})\otimes( J_{\rm{offset}}\times J_{\rm{beam}})^* S \begin{bmatrix}
         I \\
        Q \\
        U \\
        V 
    \end{bmatrix}~.
\end{split}
\end{equation*}

Applying another set of math manipulations:

\begin{equation*}
\begin{split}
   = &S^{-1} (J_{\rm{offset}}^{-1}\otimes (J_{\rm{offset}}^{-1})^*) S \\
   &\{S^{-1}  (J_{\rm{offset}}\times J_{\rm{beam}})\otimes( J_{\rm{offset}}\times J_{\rm{beam}})^* S\} \begin{bmatrix}
         I \\
        Q \\
        U \\
        V 
    \end{bmatrix}~.
\end{split}
\end{equation*}

The term within the curly braces corresponds to the Muller matrix for the beam and offset case. Thus, 


\begin{equation*}
   \begin{bmatrix}
        \mathcal{V}_I \\
        \mathcal{V}_Q \\
        \mathcal{V}_U \\
        \mathcal{V}_V 
    \end{bmatrix}_{\rm{beam}}= S^{-1}(J_{\rm{offset}}^{-1}\otimes (J_{\rm{offset}}^{-1})^*)S \begin{bmatrix}
        \mathcal{V}_I \\
        \mathcal{V}_Q \\
        \mathcal{V}_U \\
        \mathcal{V}_V 
    \end{bmatrix}_{\rm{beam+offset}}~.
\end{equation*}

Since the offset Jones matrix is diagonal and unitary (\autoref{eqn:jones-offset}), it has the property of $J_{\rm{offset}}^{-1} = J_{\rm{offset}}^*$. The correction factor in the above equation can be further simplified to:

\begin{equation*}
   \begin{bmatrix}
        \mathcal{V}_I \\
        \mathcal{V}_Q \\
        \mathcal{V}_U \\
        \mathcal{V}_V 
    \end{bmatrix}_{\rm{beam}}= S^{-1}(J_{\rm{offset}}^{*}\otimes J_{\rm{offset}})S \begin{bmatrix}
        \mathcal{V}_I \\
        \mathcal{V}_Q \\
        \mathcal{V}_U \\
        \mathcal{V}_V 
    \end{bmatrix}_{\rm{beam+offset}}~.
\end{equation*}

\begin{figure*}
\centering
\includegraphics[width=\textwidth]{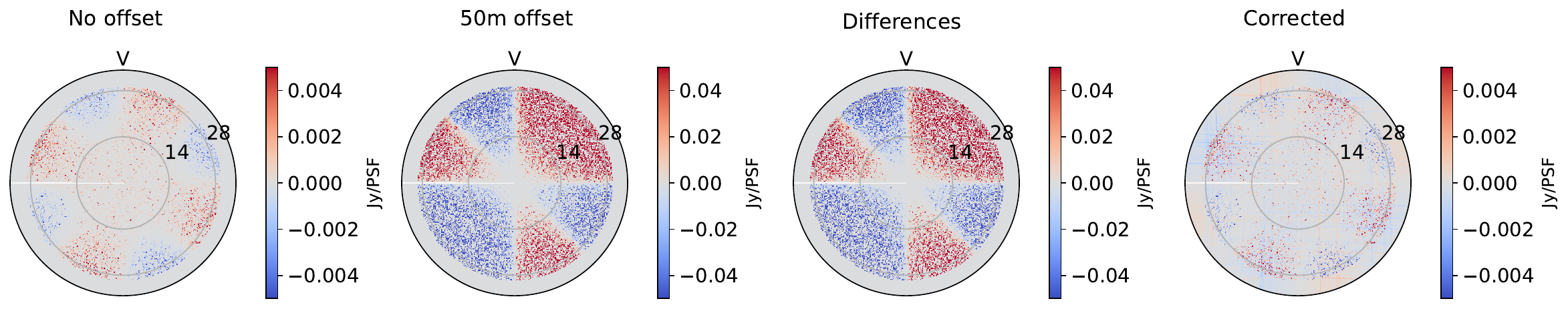}
\vspace*{-5ex}
\caption[Simulated Images of GLEAM through the Pipeline]{Similar to Fig \ref{fig:GLEAM_image} but only for Stokes V and with an ideal PSF (generated with a uniform \textit{uv}-coverage). The four image columns correspond to: co-located X and Y dipoles, with 50m offset between the X and Y dipoles, differences between co-located and non-colocated cases and offset corrected image.}
\label{fig:GLEAM_image_V}
\end{figure*}

We apply this correction on the obtained offset images (second column of \autoref{fig:GLEAM_image}) to remove the effect of dipole phase offsets and plot the resulting Stokes images in the last two columns of \autoref{fig:GLEAM_image}. In the fourth column, we have applied the correction directly to the undeconvolved images. The correction does not affect the Stokes I and Q images as expected. The correction is seen to have improved both Stokes U and V images, with the levels of Stokes U leakage being similar to the co-located case. We didn't exactly recover the polarisation levels of the non-offset cases because the correction was not applied to the deconvolved images, which will be the case when we have actual data from the array and use packages like CASA or WSClean to deconvolve the images. The residuals seen in corrected Stokes V images are due to the PSF and its sidelobes. To perform another test and show that the correction works as expected, we generate Stokes V images using the same pipeline developed, but assuming we have a perfect \textit{uv-}coverage. This is shown in figure \ref{fig:GLEAM_image_V} where each column corresponds to the Stokes V image from: No offset, with 50m offset, differences between the two cases, and corrected for offset. A perfect uv-coverage produces an ideal delta function for the PSF, implying the images generated don't require a deconvolution step. Hence, when the correction for the offset is applied, it recovers the Stokes V levels to the case without offset.

\section{Conclusion}
\label{sec:conclusion}
The Farside Array for Radio Science Investigations of the Dark Ages and Exoplanets (FARSIDE) is a NASA-funded probe-class mission concept to build a low-frequency array on the Moon. Like many other proposed lunar radio telescopes, it is a very timely mission because of all the infrastructure development as a part of NASA's Artemis program and the CLPS missions. Lunar telescopes like FARSIDE are being designed to primarily probe two unexplored low-frequency astrophysics science cases: the study of exoplanet magnetospheres and the Dark Ages epoch of the early Universe. 
The ideal location for these planned low-frequency radio arrays is the lunar farside, where there is no ionosphere and the spectra below 30MHz is shield from any terrestrial RFI. Being on the lunar farside, these radio interferometers need new strategies for antenna deployment and placement. 

In this work, we discuss the FARSIDE strategy, which will employ robotic rovers for deployment and will place dipoles directly on the regolith. We analyse and quantify the effects of this strategy on the imaging and polarisation performance of the array. To do this, we developed a custom simulation pipeline that forward-models the effects of beams, antenna positions, and array layout on all four Stokes polarisation images. The robotic deployment scheme will result in an offset between the phase centres of orthogonal dipoles of the same antenna. This is non-traditional compared to ground-based dual-polarised radio arrays, where orthogonal dipoles of the same antenna node share the same phase centre. The effects of this phase offset and the design choice of dipoles on regolith on the array's performance are studied in detail in this paper.

First, we presented a simple setup by plotting the \textit{uv}-coverages and PSFs to understand the systematics produced by non-co-located dipoles versus the traditional co-located dipoles. For a more detailed analysis of the inter-mixing of polarisations due to dipole beams and phase offsets, we use the Muller formalism. The framework of which is laid out in detail for both co-located and non-co-located dipoles. We used the Muller matrix formulation to quantify the intermixing of the various polarisation components from the sky.

For the quantitative analysis, we used the direction-dependent beams of the dipoles on the regolith, which were carefully generated with electromagnetic simulations. The simulated beams feed into the Jones matrices to calculate the Muller parameters. We also define the phase-offset Jones matrix and combine that with the beam Jones to generate the Muller parameters for the non-co-located case.

We quantify the leakages using the Intrinsic Cross-polarisation metrics (IXR). We defined two kinds of IXM that capture the relevant polarisation inter-mixing for the two interested science cases: IXR$_{M, I}$ and IXR$_{M, V}$ quantify leakage into pseudo-Stokes I and V, respectively. IXR$_{M, I}$ is only affected by the direction-dependent dipole-on-the-regolith beams and not by the phase offsets. Quantitatively, we report that the beam results in 20\%, 7.5\%, and 0.002\% leakage of the sky's Q, U, and V components into the obtained pseudo-Stokes I. For the exoplanet science case, we need high-fidelity circular polarisation data or low IXR$_{M, V}$. The beam and the phase offset affect the IXR$_{M, V}$. At approximately the centre of the FARSIDE band (2~MHz), the maximum value of the IXR$_{M, V}$ is 6\% just due to the beam, which increases by a factor of 100 due to the phase offsets. Thus, it certainly warrants a correction for the phase offsets. 

We quantified the polarisation leakages on a model sky as a final analysis. For this, we used the custom simulation pipeline we developed, the co-located and non-co-located Muller matrices, and the FARSIDE array layout. The input sky model was pure Stokes I point sources, and we made the small angle/flat sky approximation ($w=0$). We obtained similar levels of leakages as shown by Muller matrices. We looked at all four Stokes sky dirty images and reported that the maximum Stokes V/I increased from $5*10^{-5}$ for just beam effects to $5*10^{-3}$ for the beam plus offset effects, calculated at the centre of the FARSIDE band (2~MHz).

 We note that the array layout analysed in this work has all the Y-dipoles offset in the same direction with respect to their corresponding X-dipoles. This offset configuration captures the extreme case scenario in terms of polarisation leakages. However, with more careful planning in the future, offsetting the four different arms in four different directions might work out better in terms of spool length and distance travelled by the rovers. In that case, the XX and YY PSFs will not be identical, and the polarisation leakages may be lower. For the offsets in all four directions, the delta X and delta Y can take four combinations between +/-50. The offset Muller matrices (shown for $\Delta x = \Delta y = 50~m$ in Figure \ref{fig:only_offset}) will rotate by 90 degrees for $\Delta x = 50~m$ and $\Delta y = -50~m$, by $180^o$ for $\Delta x = \Delta y = -50~m$, and by $270^o$ for $\Delta x = -50~m, \Delta y = 50$. So qualitatively, it might smooth out some of the polarisation leakage differences seen in Stokes U and V of the obtained dirty images in Figure. \ref{fig:GLEAM_image}. For exact quantitative analysis, it would require adding a significant capability to the pipeline we have built for this paper, i.e., implementing antenna-dependent beams. We leave that for a future work.

The effects of the non-co-located dipoles can be corrected if the positions of each dipole are known precisely. In future work, we plan to extend the analysis to include errors in the knowledge of the dipole positions in the simulation pipeline. This will help quantify and estimate the error tolerance level needed during the FARSIDE array deployment. The same pipeline can also be used to carefully investigate tolerance levels for errors in dipole rotation and undulations due to the ground. Overall, the analysis presented here and the pipeline developed in this work can be used for forecasting any of the upcoming lunar-based radio interferometers.

\section*{ACKNOWLEDGMENTS}

The FARSIDE work was directly supported by the NASA Solar System Exploration Virtual Institute cooperative agreement 80ARC017M0006. N.M. was supported by the Future Investigators in NASA Earth and Space Science and Technology (FINESST) cooperative agreement 80NSSC19K1413.

This article is based on the author's doctoral dissertation, "Impact of Low-Frequency Antenna Characteristics on Observations of Cosmic Dawn with EDGES and FARSIDE", completed at Arizona State University.

\section*{CODE AVAILABILITY}
The code used to perform the analysis presented in this study is publicly available on GitHub at: \url{https://github.com/nmahesh1412/FARSIDE_imaging_performance}.

\bibliography{main}

\begin{thebibliography}{}
\expandafter\ifx\csname natexlab\endcsname\relax\def\natexlab#1{#1}\fi

\bibitem[{Anderson {et~al.}(2019)Anderson, Hallinan, Eastwood, Monroe, Callister, Dowell, Hicks, Huang, Kassim, Kocz, Lazio, Price, Schinzel, \& Taylor}]{Anderson2019}
Anderson, M.~M., Hallinan, G., Eastwood, M.~W., {et~al.} 2019, The Astrophysical Journal, 886, 123

\bibitem[{{Asad} {et~al.}(2016){Asad}, {Koopmans}, {Jeli{\'c}}, {Ghosh}, {Abdalla}, {Brentjens}, {de Bruyn}, {Ciardi}, {Gehlot}, {Iliev}, {Mevius}, {Pandey}, {Yatawatta}, \& {Zaroubi}}]{asad2016}
{Asad}, K.~M.~B., {Koopmans}, L.~V.~E., {Jeli{\'c}}, V., {et~al.} 2016, \mnras, 462, 4482

\bibitem[{{Bale} {et~al.}(2023){Bale}, {Bassett}, {Burns}, {Dorigo Jones}, {Goetz}, {Hellum-Bye}, {Hermann}, {Hibbard}, {Maksimovic}, {McLean}, {Monsalve}, {O'Connor}, {Parsons}, {Pulupa}, {Pund}, {Rapetti}, {Rotermund}, {Saliwanchik}, {Slosar}, {Sundkvist}, \& {Suzuki}}]{Bale_2023}
{Bale}, S.~D., {Bassett}, N., {Burns}, J.~O., {et~al.} 2023, arXiv e-prints, arXiv:2301.10345

\bibitem[{{Berger} {et~al.}(2009){Berger}, {Rutledge}, {Phan-Bao}, {Basri}, {Giampapa}, {Gizis}, {Liebert}, {Mart{\'\i}n}, \& {Fleming}}]{berger2009}
{Berger}, E., {Rutledge}, R.~E., {Phan-Bao}, N., {et~al.} 2009, \apj, 695, 310

\bibitem[{{Bernardi} {et~al.}(2013){Bernardi}, {Greenhill}, {Mitchell}, {Ord}, {Hazelton}, {Gaensler}, {de Oliveira-Costa}, {Morales}, {Udaya Shankar}, {Subrahmanyan}, {Wayth}, {Lenc}, {Williams}, {Arcus}, {Arora}, {Barnes}, {Bowman}, {Briggs}, {Bunton}, {Cappallo}, {Corey}, {Deshpande}, {deSouza}, {Emrich}, {Goeke}, {Herne}, {Hewitt}, {Johnston-Hollitt}, {Kaplan}, {Kasper}, {Kincaid}, {Koenig}, {Kratzenberg}, {Lonsdale}, {Lynch}, {McWhirter}, {Morgan}, {Oberoi}, {Pathikulangara}, {Prabu}, {Remillard}, {Rogers}, {Roshi}, {Salah}, {Sault}, {Srivani}, {Stevens}, {Tingay}, {Waterson}, {Webster}, {Whitney}, {Williams}, \& {Wyithe}}]{Bernardi2013}
{Bernardi}, G., {Greenhill}, L.~J., {Mitchell}, D.~A., {et~al.} 2013, \apj, 771, 105

\bibitem[{{Bilitza}(2018)}]{Bilitza2018}
{Bilitza}, D. 2018, Advances in Radio Science, 16, 1

\bibitem[{{Boischot} {et~al.}(1980){Boischot}, {Rosolen}, {Aubier}, {Daigne}, {Genova}, {Leblanc}, {Lecacheux}, {de La Noe}, \& {Moller-Pedersen}}]{Boischot1980}
{Boischot}, A., {Rosolen}, C., {Aubier}, M.~G., {et~al.} 1980, \icarus, 43, 399

\bibitem[{Born {et~al.}(1999)Born, Wolf, Bhatia, Clemmow, Gabor, Stokes, Taylor, Wayman, \& Wilcock}]{born_wolf_bhatia_clemmow_gabor_stokes_taylor_wayman_wilcock_1999}
Born, M., Wolf, E., Bhatia, A.~B., {et~al.} 1999, Principles of Optics: Electromagnetic Theory of Propagation, Interference and Diffraction of Light, 7th edn. (Cambridge University Press), doi:10.1017/CBO9781139644181

\bibitem[{{Braude} {et~al.}(1978){Braude}, {Megn}, {Rashkovskii}, {Riabov}, {Sharykin}, {Sokolov}, {Tkachenko}, \& {Zhuk}}]{Braude1978}
{Braude}, S.~I., {Megn}, A.~V., {Rashkovskii}, S.~L., {et~al.} 1978, \apss, 54, 37

\bibitem[{{Bridle}(1967)}]{bridle1967}
{Bridle}, A.~H. 1967, \mnras, 136, 219

\bibitem[{{Burke} \& {Franklin}(1955)}]{Burke1955}
{Burke}, B.~F., \& {Franklin}, K.~L. 1955, \jgr, 60, 213

\bibitem[{Burns {et~al.}(2021{\natexlab{a}})Burns, Hallinan, Chang, Anderson, Bowman, Bradley, Furlanetto, Hegedus, Kasper, Kocz, Lazio, Lux, MacDowall, Mirocha, Nesnas, Pober, Polidan, Rapetti, Romero-Wolf, Slosar, Stebbins, Teitelbaum, \& White}]{burns2021lunar}
Burns, J., Hallinan, G., Chang, T.-C., {et~al.} 2021{\natexlab{a}}, A Lunar Farside Low Radio Frequency Array for Dark Ages 21-cm Cosmology, arXiv:2103.08623

\bibitem[{Burns {et~al.}(2021{\natexlab{b}})Burns, MacDowall, Bale, Hallinan, Bassett, \& Hegedus}]{burns2021nasa}
Burns, J.~O., MacDowall, R., Bale, S., {et~al.} 2021{\natexlab{b}}, Low Radio Frequency Observations from the Moon Enabled by NASA Landed Payload Missions, arXiv:2102.02331

\bibitem[{Burns {et~al.}(2019a)Burns, Hallinan, Teitelbaum, Chang, Kocz, Bowman, MacDowall, Kasper, Bradley, Anderson, Rapetti, Zhen, Wu, Pober, Furlanetto, Mirocha, \& Austin}]{Burns2019a}
Burns, J.~O., Hallinan, G., Teitelbaum, L., {et~al.} 2019a, “Probe Study Report: FARSIDE (Farside Array for Radio Science Investigations of the Darkages and Exoplanets”, NASA

\bibitem[{Byrne {et~al.}(2021)Byrne, Morales, Hazelton, Sullivan, Barry, Lynch, Line, \& Jacobs}]{Bryne_2022}
Byrne, R., Morales, M.~F., Hazelton, B., {et~al.} 2021, Monthly Notices of the Royal Astronomical Society, 510, 2011

\bibitem[{{Cane}(1978)}]{Cane1978}
{Cane}, H.~V. 1978, Australian Journal of Physics, 31, 561

\bibitem[{Carozzi \& Woan(2011)}]{carozzi2011}
Carozzi, T.~D., \& Woan, G. 2011, IEEE Transactions on Antennas and Propagation, 59, 2058

\bibitem[{Chen {et~al.}(2019)Chen, Burns, Koopmans, Rothkaehi, Silk, Wu, Boonstra, Cecconi, Chiang, Chen, Deng, Falanga, Falcke, Fan, Fang, Fialkov, Gurvits, Ji, Kasper, Li, Mao, Mckinley, Monsalve, Peterson, Ping, Subrahmanyan, Vedantham, Wolt, Wu, Xu, Yan, \& Yue}]{chen2019}
Chen, X., Burns, J., Koopmans, L., {et~al.} 2019, Discovering the Sky at the Longest Wavelengths with Small Satellite Constellations, arXiv:1907.10853

\bibitem[{{Chiang} {et~al.}(2020){Chiang}, {Dyson}, {Egan}, {Eyono}, {Ghazi}, {Hickish}, {J{\'a}uregui-Garcia}, {Manukha}, {M{\'e}nard}, {Moso}, {Peterson}, {Philip}, {Sievers}, \& {Tartakovsky}}]{Chiang_2020}
{Chiang}, H.~C., {Dyson}, T., {Egan}, E., {et~al.} 2020, Journal of Astronomical Instrumentation, 9, 2050019

\bibitem[{Clark {et~al.}(2018)Clark, Dutta, Gao, Ma, \& Strigari}]{Clark2018}
Clark, S.~J., Dutta, B., Gao, Y., Ma, Y.-Z., \& Strigari, L.~E. 2018, Phys. Rev. D, 98, 043006

\bibitem[{{Cosmic Visions 21 cm Collaboration} {et~al.}(2018){Cosmic Visions 21 cm Collaboration}, {Ansari}, {Arena}, {Bandura}, {Bull}, {Castorina}, {Chang}, {Chen}, {Connor}, {Foreman}, {Frisch}, {Green}, {Johnson}, {Karagiannis}, {Liu}, {Masui}, {Meerburg}, {M{\"u}nchmeyer}, {Newburgh}, {Obuljen}, {O'Connor}, {Padmanabhan}, {Shaw}, {Sheehy}, {Slosar}, {Smith}, {Stankus}, {Stebbins}, {Timbie}, {Villaescusa-Navarro}, {Wallisch}, \& {White}}]{2018ansari}
{Cosmic Visions 21 cm Collaboration}, {Ansari}, R., {Arena}, E.~J., {et~al.} 2018, arXiv e-prints, arXiv:1810.09572

\bibitem[{{Ergun} {et~al.}(2000){Ergun}, {Carlson}, {McFadden}, {Delory}, {Strangeway}, \& {Pritchett}}]{ergun2000}
{Ergun}, R.~E., {Carlson}, C.~W., {McFadden}, J.~P., {et~al.} 2000, \apj, 538, 456

\bibitem[{{Erickson} {et~al.}(1982){Erickson}, {Mahoney}, \& {Erb}}]{Erickson1982}
{Erickson}, W.~C., {Mahoney}, M.~J., \& {Erb}, K. 1982, \apjs, 50, 403

\bibitem[{Falcke(2018)}]{change}
Falcke, H. 2018, Change

\bibitem[{Farnes {et~al.}(2014)Farnes, Gaensler, \& Carretti}]{Farnes2014}
Farnes, J.~S., Gaensler, B.~M., \& Carretti, E. 2014, The Astrophysical Journal Supplement Series, 212, 15

\bibitem[{{Farrell} {et~al.}(1998){Farrell}, {Kaiser}, {Steinberg}, \& {Bale}}]{farrell1998}
{Farrell}, W.~M., {Kaiser}, M.~L., {Steinberg}, J.~T., \& {Bale}, S.~D. 1998, \jgr, 103, 23653

\bibitem[{{Gehlot} {et~al.}(2018){Gehlot}, {Koopmans}, {de Bruyn}, {Zaroubi}, {Brentjens}, {Asad}, {Hatef}, {Jeli{\'c}}, {Mevius}, {Offringa}, {Pandey}, \& {Yatawatta}}]{gehlot2018}
{Gehlot}, B.~K., {Koopmans}, L.~V.~E., {de Bruyn}, A.~G., {et~al.} 2018, \mnras, 478, 1484

\bibitem[{{Geil} {et~al.}(2011){Geil}, {Gaensler}, \& {Wyithe}}]{geil2011}
{Geil}, P.~M., {Gaensler}, B.~M., \& {Wyithe}, J. S.~B. 2011, \mnras, 418, 516

\bibitem[{{Hallinan} {et~al.}(2008){Hallinan}, {Antonova}, {Doyle}, {Bourke}, {Lane}, \& {Golden}}]{hallinan2008}
{Hallinan}, G., {Antonova}, A., {Doyle}, J.~G., {et~al.} 2008, \apj, 684, 644

\bibitem[{{Hallinan} {et~al.}(2007){Hallinan}, {Bourke}, {Lane}, {Antonova}, {Zavala}, {Brisken}, {Boyle}, {Vrba}, {Doyle}, \& {Golden}}]{hallinan2007}
{Hallinan}, G., {Bourke}, S., {Lane}, C., {et~al.} 2007, \apjl, 663, L25

\bibitem[{{Hamaker} {et~al.}(1996){Hamaker}, {Bregman}, \& {Sault}}]{hamaker1996}
{Hamaker}, J.~P., {Bregman}, J.~D., \& {Sault}, R.~J. 1996, \aaps, 117, 137

\bibitem[{{Heiken} {et~al.}(1991){Heiken}, {Vaniman}, \& {French}}]{Heiken1991}
{Heiken}, G.~H., {Vaniman}, D.~T., \& {French}, B.~M. 1991, {Lunar Sourcebook, A User's Guide to the Moon}

\bibitem[{Hill \& Baxter(2018)}]{Hill2018}
Hill, J.~C., \& Baxter, E.~J. 2018, Journal of Cosmology and Astroparticle Physics, 2018, 037

\bibitem[{{Hurley-Walker} {et~al.}(2017){Hurley-Walker}, {Callingham}, {Hancock}, {Franzen}, {Hindson}, {Kapi{\'n}ska}, {Morgan}, {Offringa}, {Wayth}, {Wu}, {Zheng}, {Murphy}, {Bell}, {Dwarakanath}, {For}, {Gaensler}, {Johnston-Hollitt}, {Lenc}, {Procopio}, {Staveley-Smith}, {Ekers}, {Bowman}, {Briggs}, {Cappallo}, {Deshpande}, {Greenhill}, {Hazelton}, {Kaplan}, {Lonsdale}, {McWhirter}, {Mitchell}, {Morales}, {Morgan}, {Oberoi}, {Ord}, {Prabu}, {Shankar}, {Srivani}, {Subrahmanyan}, {Tingay}, {Webster}, {Williams}, \& {Williams}}]{hurleywalker2017}
{Hurley-Walker}, N., {Callingham}, J.~R., {Hancock}, P.~J., {et~al.} 2017, \mnras, 464, 1146

\bibitem[{Kasper {et~al.}(2019)Kasper, Lazio, Romero-Wolf, Lux, \& Neilsen}]{Kasper2018}
Kasper, J., Lazio, J., Romero-Wolf, A., Lux, J., \& Neilsen, T. 2019, in 2019 IEEE Aerospace Conference, 1--11

\bibitem[{{Kohn} \& et. al.(2019)}]{kohn2019}
{Kohn}, S.~A., \& et. al. 2019, \apj, 882, 58

\bibitem[{{Konovalenko} {et~al.}(2016){Konovalenko}, {Sodin}, {Zakharenko}, {Zarka}, {Ulyanov}, {Sidorchuk}, {Stepkin}, {Tokarsky}, {Melnik}, {Kalinichenko}, {Stanislavsky}, {Koliadin}, {Shepelev}, {Dorovskyy}, {Ryabov}, {Koval}, {Bubnov}, {Yerin}, {Gridin}, {Kulishenko}, {Reznichenko}, {Bortsov}, {Lisachenko}, {Reznik}, {Kvasov}, {Mukha}, {Litvinenko}, {Khristenko}, {Shevchenko}, {Shevchenko}, {Belov}, {Rudavin}, {Vasylieva}, {Miroshnichenko}, {Vasilenko}, {Olyak}, {Mylostna}, {Skoryk}, {Shevtsova}, {Plakhov}, {Kravtsov}, {Volvach}, {Lytvinenko}, {Shevchuk}, {Zhouk}, {Bovkun}, {Antonov}, {Vavriv}, {Vinogradov}, {Kozhin}, {Kravtsov}, {Bulakh}, {Kuzin}, {Vasilyev}, {Brazhenko}, {Vashchishin}, {Pylaev}, {Koshovyy}, {Lozinsky}, {Ivantyshin}, {Rucker}, {Panchenko}, {Fischer}, {Lecacheux}, {Denis}, {Coffre}, {Grie{\ss}meier}, {Tagger}, {Girard}, {Charrier}, {Briand}, \& {Mann}}]{konovalenko2016}
{Konovalenko}, A., {Sodin}, L., {Zakharenko}, V., {et~al.} 2016, Experimental Astronomy, 42, 11

\bibitem[{{Koopmans} {et~al.}(2021){Koopmans}, {Barkana}, {Bentum}, {Bernardi}, {Boonstra}, {Bowman}, {Burns}, {Chen}, {Datta}, {Falcke}, {Fialkov}, {Gehlot}, {Gurvits}, {Jeli{\'c}}, {Klein-Wolt}, {Lazio}, {Meerburg}, {Mellema}, {Mertens}, {Mesinger}, {Offringa}, {Pritchard}, {Semelin}, {Subrahmanyan}, {Silk}, {Trott}, {Vedantham}, {Verde}, {Zaroubi}, \& {Zarka}}]{koopmans2021}
{Koopmans}, L. V.~E., {Barkana}, R., {Bentum}, M., {et~al.} 2021, Experimental Astronomy, 51, 1641

\bibitem[{{Lamy} {et~al.}(2017){Lamy}, {Zarka}, {Cecconi}, {Klein}, {Masson}, {Denis}, {Coffre}, \& {Viou}}]{lamy2018}
{Lamy}, L., {Zarka}, P., {Cecconi}, B., {et~al.} 2017, in Planetary Radio Emissions VIII, ed. G.~{Fischer}, G.~{Mann}, M.~{Panchenko}, \& P.~{Zarka}, 455--466

\bibitem[{{Lenc} {et~al.}(2016){Lenc}, {Gaensler}, {Sun}, {Sadler}, {Willis}, {Barry}, {Beardsley}, {Bell}, {Bernardi}, {Bowman}, {Briggs}, {Callingham}, {Cappallo}, {Carroll}, {Corey}, {de Oliveira-Costa}, {Deshpande}, {Dillon}, {Dwarkanath}, {Emrich}, {Ewall-Wice}, {Feng}, {For}, {Goeke}, {Greenhill}, {Hancock}, {Hazelton}, {Hewitt}, {Hindson}, {Hurley-Walker}, {Johnston-Hollitt}, {Jacobs}, {Kapi{\'n}ska}, {Kaplan}, {Kasper}, {Kim}, {Kratzenberg}, {Line}, {Loeb}, {Lonsdale}, {Lynch}, {McKinley}, {McWhirter}, {Mitchell}, {Morales}, {Morgan}, {Morgan}, {Murphy}, {Neben}, {Oberoi}, {Offringa}, {Ord}, {Paul}, {Pindor}, {Pober}, {Prabu}, {Procopio}, {Riding}, {Rogers}, {Roshi}, {Udaya Shankar}, {Sethi}, {Srivani}, {Staveley-Smith}, {Subrahmanyan}, {Sullivan}, {Tegmark}, {Thyagarajan}, {Tingay}, {Trott}, {Waterson}, {Wayth}, {Webster}, {Whitney}, {Williams}, {Williams}, {Wu}, {Wyithe}, \& {Zheng}}]{Lenc_2016}
{Lenc}, E., {Gaensler}, B.~M., {Sun}, X.~H., {et~al.} 2016, \apj, 830, 38

\bibitem[{Liu {et~al.}(2011)Liu, Chen, Le, Kurkin, Polekh, \& Lee}]{liu2011}
Liu, L., Chen, Y., Le, H., {et~al.} 2011, Journal of Geophysical Research: Space Physics, 116, https://agupubs.onlinelibrary.wiley.com/doi/pdf/10.1029/2010JA016296

\bibitem[{Mao {et~al.}(2008)Mao, Tegmark, McQuinn, Zaldarriaga, \& Zahn}]{Mao2008}
Mao, Y., Tegmark, M., McQuinn, M., Zaldarriaga, M., \& Zahn, O. 2008, Physical Review D, 78, doi:10.1103/physrevd.78.023529

\bibitem[{{McGarey} {et~al.}(2022){McGarey}, {Nesnas}, {Rajguru}, {Bezkrovny}, {Jamnejad}, {Lux}, {Sunada}, {Teitelbaum}, {Miller}, {Squyres}, {Hallinan}, {Hegedus}, \& {Burns}}]{McGarey2022}
{McGarey}, P., {Nesnas}, I.~A., {Rajguru}, A., {et~al.} 2022, arXiv e-prints, arXiv:2209.02216

\bibitem[{Muñoz {et~al.}(2015)Muñoz, Kovetz, \& Ali-Haïmoud}]{munoz2015}
Muñoz, J.~B., Kovetz, E.~D., \& Ali-Haïmoud, Y. 2015, Physical Review D, 92, doi:10.1103/physrevd.92.083528

\bibitem[{Nesnas {et~al.}(2012)Nesnas, Matthews, Abad-Manterola, Burdick, Edlund, Morrison, Peters, Tanner, Miyake, Solish, \& Anderson}]{nesnas2012}
Nesnas, I.~A., Matthews, J.~B., Abad-Manterola, P., {et~al.} 2012, Journal of Field Robotics, 29, 663

\bibitem[{{Nunhokee} {et~al.}(2017){Nunhokee}, {Bernardi}, {Kohn}, {Aguirre}, {Thyagarajan}, {Dillon}, {Foster}, {Grobler}, {Martinot}, \& {Parsons}}]{nunhokee2017}
{Nunhokee}, C.~D., {Bernardi}, G., {Kohn}, S.~A., {et~al.} 2017, \apj, 848, 47

\bibitem[{{O'Sullivan} {et~al.}(2023){O'Sullivan}, {Shimwell}, {Hardcastle}, {Tasse}, {Heald}, {Carretti}, {Br{\"u}ggen}, {Vacca}, {Sobey}, {Van Eck}, {Horellou}, {Beck}, {Bilicki}, {Bourke}, {Botteon}, {Croston}, {Drabent}, {Duncan}, {Heesen}, {Ideguchi}, {Kirwan}, {Lawlor}, {Mingo}, {Nikiel-Wroczy{\'n}ski}, {Piotrowska}, {Scaife}, \& {van Weeren}}]{sullivan_2023}
{O'Sullivan}, S.~P., {Shimwell}, T.~W., {Hardcastle}, M.~J., {et~al.} 2023, \mnras, 519, 5723

\bibitem[{{Patsourakos} \& {Georgoulis}(2017)}]{patsourakos2017}
{Patsourakos}, S., \& {Georgoulis}, M.~K. 2017, \solphys, 292, 89

\bibitem[{Pritchard \& Loeb(2012)}]{loeb2012}
Pritchard, J.~R., \& Loeb, A. 2012, RPP, 75, 086901

\bibitem[{{Reber}(1994)}]{Reber1994}
{Reber}, G. 1994, \jrasc, 88, 297

\bibitem[{{Roger} {et~al.}(1999){Roger}, {Costain}, {Landecker}, \& {Swerdlyk}}]{Roger1999}
{Roger}, R.~S., {Costain}, C.~H., {Landecker}, T.~L., \& {Swerdlyk}, C.~M. 1999, \aaps, 137, 7

\bibitem[{{Rogers} {et~al.}(2015){Rogers}, {Bowman}, {Vierinen}, {Monsalve}, \& {Mozdzen}}]{Rogers2015}
{Rogers}, A.~E.~E., {Bowman}, J.~D., {Vierinen}, J., {Monsalve}, R., \& {Mozdzen}, T. 2015, Radio Science, 50, 130

\bibitem[{Slatyer(2013)}]{Slatyer2013}
Slatyer, T.~R. 2013, Physical Review D, 87, doi:10.1103/physrevd.87.123513

\bibitem[{Slatyer(2016)}]{slayter2016}
---. 2016, Physical Review D, 93, doi:10.1103/physrevd.93.023527

\bibitem[{{Smirnov}(2011)}]{smirnov2011}
{Smirnov}, O.~M. 2011, \aap, 527, A106

\bibitem[{{Thompson} {et~al.}(2001){Thompson}, {Moran}, \& {Swenson}}]{thompson2001}
{Thompson}, A.~R., {Moran}, J.~M., \& {Swenson}, George~W., J. 2001, {Interferometry and Synthesis in Radio Astronomy, 2nd Edition}

\bibitem[{{Van Eck} {et~al.}(2018){Van Eck}, {Haverkorn}, {Alves}, {Beck}, {Best}, {Carretti}, {Chy{\.z}y}, {Farnes}, {Ferri{\`e}re}, {Hardcastle}, {Heald}, {Horellou}, {Iacobelli}, {Jeli{\'c}}, {Mulcahy}, {O'Sullivan}, {Polderman}, {Reich}, {Riseley}, {R{\"o}ttgering}, {Schnitzeler}, {Shimwell}, {Vacca}, {Vink}, \& {White}}]{Eck_2018}
{Van Eck}, C.~L., {Haverkorn}, M., {Alves}, M.~I.~R., {et~al.} 2018, \aap, 613, A58

\bibitem[{{Vecchio} {et~al.}(2021){Vecchio}, {Bentum}, {Falcke}, {Boonstra}, {Ping}, {Chen}, {Klein-Wolt}, {Brinkerink}, {Rotteveel}, {Pourshaghaghi}, {Karapakula}, {Ruiter}, \& {Bertels}}]{Vecchio2021}
{Vecchio}, A., {Bentum}, M., {Falcke}, H., {et~al.} 2021, in 43rd COSPAR Scientific Assembly. Held 28 January - 4 February, Vol.~43, 1525

\bibitem[{Vedantham \& Koopmans(2015)}]{vendantham2015}
Vedantham, H.~K., \& Koopmans, L. V.~E. 2015, Monthly Notices of the Royal Astronomical Society, 453, 925–938

\bibitem[{{Zarka}(1998)}]{zarka1998}
{Zarka}, P. 1998, \jgr, 103, 20159

\bibitem[{{Zarka} {et~al.}(2004){Zarka}, {Cecconi}, \& {Kurth}}]{zarka2004}
{Zarka}, P., {Cecconi}, B., \& {Kurth}, W.~S. 2004, Journal of Geophysical Research (Space Physics), 109, A09S15

\bibitem[{{Zhao} {et~al.}(2019){Zhao}, {Liu}, {Zhang}, {Zhou}, {Yang}, {He}, {Gao}, \& {Xiao}}]{zhao2019}
{Zhao}, W., {Liu}, S., {Zhang}, S., {et~al.} 2019, \grl, 46, 7230

\end{thebibliography}

\end{document}